\documentclass[journal,onecolumn]{IEEEtran}
\usepackage[colorlinks, linkcolor=red, citecolor=blue]{hyperref}
\usepackage{amsfonts,color,morefloats,pslatex}
\usepackage{amsmath}
\usepackage{amssymb,amsthm}
\usepackage{latexsym}
\usepackage{cite}
\usepackage{enumerate}
\usepackage{lineno,hyperref,mathtools}
\usepackage[ruled,linesnumbered]{algorithm2e}
\usepackage{pdflscape}
\usepackage{setspace}
\usepackage{pifont}
\usepackage{booktabs}
\allowdisplaybreaks
\newtheorem{theorem}{Theorem}

\newtheorem{lemma}{Lemma}
\newtheorem{corollary}{Corollary}

\theoremstyle{definition}
\newtheorem{definition}{Definition}
\newtheorem{example}{Example}

\newtheorem{problem}{Problem}
\newtheorem{remark}[theorem]{Remark}

\usepackage{soul}

\newcommand{\F}{\mathbb{F}}

\newcommand{\ccc}{{\mathbf{c}}}
\newcommand{\vvv}{{\mathbf{v}}}
\newcommand{\aaa}{{\mathbf{a}}}
\newcommand{\C}{{\mathcal{C}}}
\newcommand{\D}{{\mathcal{D}}}
\newcommand{\supp}{{\rm{supp}}}
\newcommand{\SSS}{{\mathcal{S}}}

\newcommand{\GRS}{{\mathrm{GRS}}}
\newcommand{\EGRS}{{\mathrm{EGRS}}}
\newcommand{\TGRS}{{\mathrm{TGRS}}}
\newcommand{\ETGRS}{{\mathrm{ETGRS}}}
\newcommand{\diag}{{\mathrm{diag}}}

\newcommand{\uuu}{{{\mathbf{u}}}}
\newcommand{\www}{{{\mathbf{w}}}}
\newcommand{\rrr}{{{\mathbf{r}}}}
\newcommand{\w}{{{\omega}}}
\newcommand{\wt}{{{\rm{wt}}}}

\makeatletter

\newcommand{\Rmnum}[1]{\expandafter\@slowromancap\romannumeral #1@}
\makeatother
\makeatletter
\renewcommand*\env@matrix[1][\arraystretch]{%
  \edef\arraystretch{#1}%
  \hskip -\arraycolsep
  \let\@ifnextchar\new@ifnextchar
  \array{*\c@MaxMatrixCols c}}
\makeatother

\begin{document}
%
\title{Properties and Decoding of Twisted GRS Codes and Their Extensions}
\author{Yang Li, Martianus Frederic Ezerman, Huimin Lao, San Ling 
\thanks{Yang Li, Martianus Frederic Ezerman, Huimin Lao, and San Ling are with the School of Physical and Mathematical Sciences, 
Nanyang Technological University, 21 Nanyang Link, 637371, Singapore.
(Emails: li-y@ntu.edu.sg, fredezerman@ntu.edu.sg, huimin.lao@ntu.edu.sg, and lingsan@ntu.edu.sg.)}
\thanks{
This research is supported by the Nanyang Technological University Research under Grant 04INS000047C230GRT01 and the Anhui Provincial Natural Science Foundation Grant 2408085MA014.} 
}

\maketitle

\begin{abstract}
Maximum distance separable (MDS) codes that are not equivalent to generalized Reed-Solomon (GRS) codes are called non-GRS MDS codes. Alongside near MDS (NMDS) codes, they are applicable in communication, cryptography, and storage systems. From theoretical perspective, it is particularly intriguing to investigate families of linear codes in which each element can be determined to be either a non-GRS MDS or an NMDS code. Two promising candidates for such families emerge from what is known as twisted GRS (TGRS) construction. These candidates are the $(+)$-TGRS codes and their extended versions, called $(+)$-extended TGRS (ETGRS) codes. 

Although many of their properties have been characterized, there are gaps to fill. Which among the codes are non-GRS MDS? Can we improve on their decoding by using their error-correcting pairs or deep holes? In this paper we solve these problems. The answer to the first problem leads us to two classes of non-GRS MDS Hermitian self-dual TGRS codes and a proof that there is no Galois self-dual ETGRS code. Addressing the second problem, we present an explicit decoding algorithm for ETGRS codes that outperforms existing decoding algorithms given some conditions. By considering the duals of TGRS codes which are MDS, we determine the covering radius and a class of deep holes of the recently constructed non-GRS MDS codes due to Han and Zhang.
\end{abstract}



\section{Introduction}\label{sec.introduction}

Let $\F_q$ be the finite field of size $q=p^m$ for some prime $p$ and positive integer $m$. Let $\F_q^n$ be the $n$-dimensional vector space over $\F_q$. 
An $[n,k,d]_q$ {\em linear code} $\C$ is a $k$-dimensional subspace of $\F_q^n$ whose minimum distance is $d$. The parameters of $\C$ satisfy the well-known {\em Singleton bound} $d\leq n-k+1$ treated in, {\it e.g.}, \cite{HP2003}.
We say that $\C$ is a {\em maximum distance separable} (MDS) code if $d=n-k+1$. The code is {\em almost MDS} (AMDS) if $d=n-k$. If {\em both} $\C$ and its dual $\C^{\perp}$ are AMDS under a suitable non-degenerate inner product, then $\C$ is {\em near MDS} (NMDS). Both MDS and NMDS codes have been considered for applications in distributed storage systems \cite{CHL2011}. Their deep connections to other objects have also been confirmed. Prominent examples include their structural links to finite geometries in \cite{DL1994}, random error channels in \cite{ST2013}, secret sharing schemes in \cite{SV2018,ZWXLQY2009}, and informed source and index coding problems in \cite{TR2018}.   

\medskip
\noindent
{\bf GRS or Non-GRS}
\smallskip

Generalized Reed–Solomon (GRS) codes form a special class of MDS codes. Codes which are MDS but not equivalent to GRS codes are called {\em non-GRS MDS codes}. The study of the latter is fundamental in the classification of MDS codes \cite{PD1991,C2024,LCCN2025,JMXZ2024}. 
In code-based cryptography, the McEliece cryptosystem faces significant challenges in gaining traction due to its huge public key sizes inherited from the underlying family of binary Goppa codes. Other families of codes, including non-GRS MDS codes have been studied in \cite{BBPR2018} and \cite{LR2020} as alternatives. The hope is to significantly reduce the key sizes while offering at least the same resistance to cryptanalysis, both classical and quantum.

The sustained interest from both theoretical and practical perspectives, provides a strong motivation to study the properties of new families of linear codes in which each code is either a non-GRS MDS code or an NMDS code. Two promising candidates for such families are $(+)$-twisted GRS codes and $(+)$-extended TGRS codes. They are often abbreviated to $(+)$-TGRS and $(+)$-ETGRS codes, respectively.

\begin{definition}\label{def.ETGRS code}
For fixed $q$ and $n$, let $k$ be such that $3\leq k\leq n-2\leq q-2$. Let $\SSS := \{a_1,a_2,\ldots,a_n\} \subseteq \F_q$. Let $\vvv :=(v_1,v_2,\ldots,v_n)\in (\F_q^*)^n$ and let $\eta\in \F_q^*$. Let  
\begin{equation}\label{eq:space}
\mathbb{V}_{q}[x]_{k} : = \left\{f(x)= \sum_{i=0}^{k-1}f_i \, x^i + \eta \, f_{k-1} \, x^{k} \, : \, f_i\in \F_q \mbox{ for all } 0\leq i\leq k-1\right\}, 
\end{equation}
with $f_{k-1}$ being the coefficient of $x^{k-1}$ in $f(x)$. An $[n,k]_{q}$ {\em $(+)$-twisted generalized Reed-Solomon code} associated with $\SSS$ and $\mathbf{v}$ is defined by
\begin{equation}\label{eq.TGRS express}
\TGRS_k(\SSS,\mathbf{v},\eta) =  \left\{(v_1f(a_1),v_2f(a_2),\ldots,v_nf(a_n)) \, : \,  f(x)\in \mathbb{V}_{q}[x]_{k}\right\}.
\end{equation} 
An $[n+1,k]_{q}$ {\em $(+)$-extended TGRS code} associated with $\SSS$ and $\mathbf{v}$ is defined by 
\begin{equation}\label{eq.ETGRS express}
\ETGRS_k(\SSS,\mathbf{v},\eta,\infty) = 
\left\{(v_1f(a_1),v_2f(a_2),\ldots,v_nf(a_n),f_{k-1}) \, : \,  f(x)\in \mathbb{V}_{q}[x]_{k}\right\}.
\end{equation}
\end{definition}

Huang \textit{et al.} in \cite{HYN2021DCC} as well as Zhu and Liao in \cite{ZL2024} have shown that any $(+)$-TGRS or $(+)$-ETGRS code $\C$ must be either MDS or NMDS. They proved necessary and sufficient conditions to determine which of the two possibilities holds given such a code $\C$. Beelen \textit{et al.} established in \cite{BBPR2018} that any $(+)$-TGRS code $\C$ which is MDS must be non-GRS whenever $3\leq k<\frac{n}{2}$. He and Liao in \cite{HL2023DM} extended the range of $k$ to $\frac{n}{2}<k\leq n-3$ while maintaining the finding of Beelen {\it et al}. In their already cited work above, Zhu and Liao claimed that \emph{all} $(+)$-ETGRS codes are non-GRS without proving it for the case of $k=\frac{n+1}{2}$ for odd $n$. We know from the work of Jin \textit{et al.} in \cite{JMXZ2024} that any $[n,k,n-k+1]_q$ MDS code must be GRS whenever $\min\{k,n-k\}<3$. In summary, we have the following problems to settle before we can close the remaining gaps.
\begin{problem}\label{prob.1} {\em (Deciding if a $(+)$-TGRS or $(+)$-ETGRS MDS code is non-GRS)}
\begin{enumerate}[1.]
\item For a given even $n$, are all $(+)$-TGRS codes with parameters 
$\displaystyle{\left[n,\frac{n}{2},\frac{n}{2}+1\right]_q}$ non-GRS?
\item For a given odd $n$, are all $(+)$-ETGRS codes with parameters $\displaystyle{\left[n+1,\frac{n+1}{2},\frac{n+1}{2}+1\right]_q}$ non-GRS?
\end{enumerate}
\end{problem}

\medskip
\noindent
{\bf Decoding}
\smallskip

Having codes with good parameters usually triggers us to find significant applications and efficient decoding \cite{HP2003,LX2004}. There are primarily two approaches to decode $(+)$-TGRS and $(+)$-ETGRS codes. The first approach is via the $t$-error-correcting pair (ECP) whereas the second relies on the maximum likelihood method. The former leads to a unique algebraic decoding procedure capable of correcting up to $t$ errors in polynomial time as already confirmed in \cite{HL2023DM} and \cite{P1992}. The latter is closely connected to the study of deep holes, which are extreme cases that constitute the most challenging instances of maximum likelihood decoding method. More details on these cases can be found in, {\it e.g.}, \cite{FXZ2025} and \cite{ZCL2016}.  
We note, without elaborating further, that Pellikaan and Marquez-Corbella have explored the application of ECPs in quantum-resistant cryptography in \cite{PM2017}. 
In another direction, Li \textit{et al.} in \cite{LZS2025} as well as Yang and Qiu in \cite{YQ2023} used the deep holes of well-chosen codes to obtain new optimal linear codes. 

Let $\C$ be a given $(+)$-TGRS or $(+)$-ETGRS code with minimum distance $d(\C)$ and let $t:=\lfloor \frac{d(\C)-1}{2}\rfloor$. Table \ref{table:summary} summarizes known results on the $t$-ECPs.

\begin{table*}[!ht]
\caption{Known results on the $t$-ECPs of $(+)$-TGRS and $(+)$-ETGRS codes.}\label{table:summary}
\renewcommand{\arraystretch}{1.1}
\centering
\begin{tabular}{lllll}
\toprule
No. & Done by & Ref. & Results & Conditions \\
\midrule
1 & He and Liao & \cite{HL2023DM} & There is no $[n,k,n-k+1]_q$ $(+)$-TGRS code with $t$-ECP & $k\in [3,\frac{n}{2})\cup (\frac{n}{2},n-3]$\\
&& & & and $(n-k)$ is even.\\
2 & & & All $[n,k,n-k+1]_q$ $(+)$-TGRS codes have $(t-1)$-ECPs & $3\leq k\leq n-3$ and\\ && & & $(n-k)$ is even.\\
3 & & &  All $[n,k,n-k+1]_q$ $(+)$-TGRS codes have $t$-ECPs & $3\leq k\leq n-3$ \\ && & & and $(n-k)$ is odd.\\
4 & & &  All $[n,k,n-k]_q$ $(+)$-TGRS codes have $t$-ECPs & $3\leq k\leq n-3$ \\
\midrule
5 & Li, Zhu, and Sun & \cite{LZS2025} & There is no $[n+1,k,n-k+2]_q$ $(+)$-ETGRS code with $t$-ECP & $3\leq k\leq n-2$ \\
&&&& and $(n-k)$ is odd.\\
\bottomrule
\end{tabular}
\end{table*}
In their concluding remarks, the authors of \cite{LZS2025} conjectured that there exist $[n+1, k, n - k + 2]_q$ $(+)$-ETGRS MDS codes with $(t - 1)$-ECPs when $3\leq k\leq n-2$ and $(n-k)$ is odd. For other feasible values of $k$, they conjectured that the corresponding codes have $t$-ECPs. Taking into account the results obtained in \cite{HL2023DM}, the existence of $t$-ECPs for $[n+1, k, n - k + 1]_q$ $(+)$-ETGRS NMDS codes remains unknown. Hence, we have the following open problem. 
\begin{problem}\label{prob.2}{\em (Existence of $(+)$-TGRS and $(+)$-ETGRS codes with specific $t$-ECPs)}
\begin{enumerate}[1.]
\item If $n$ is divisible by $4$, do all $(+)$-TGRS codes with parameters 
$\displaystyle{\left[n,\frac{n}{2},\frac{n}{2}+1 \right]_q}$ have $t$-ECPs?
        
\item If $n-k$ is odd, do all $(+)$-ETGRS codes with parameters $[n+1,k,n-k+2]_q$ have 
$(t-1)$-ECPs? 
\item Do all $(+)$-ETGRS codes with parameters $[n+1,k,n-k+2]_q$ or $[n+1,k,n-k+1]_q$ have $t$-ECPs? For the former, $n-k$ must be even.
\end{enumerate}
\end{problem}

Fang \textit{et al.} in \cite{FXZ2025} and Li \textit{et al.} in~\cite{LZS2025} have determined the covering radius of $(+)$-TGRS and $(+)$-ETGRS codes and explored several classes of their deep holes. 
Sun \textit{et al.} in~\cite{SDC2024FFA} introduced what they called \emph{second kind of extended codes}, which we call them {\it second extended codes} here. 
If $\C$ is a given MDS code, then its second extended codes being MDS or not is closely connected to the covering radius and deep holes of $\C^{\perp}$, as it stated in \cite{WDC2023}. 
By representing an $[n+1,k,n-k+2]_q$ MDS code as a second extended code of a certain $[n,k,n-k+1]_q$ MDS code, 
Abdukhalikov \textit{et al.} in \cite{ADV2025}, Wu \textit{et al.} in \cite{WDC2023}, and Wu with a different set of coauthors in \cite{WHLD2024} further obtained the covering radius and deep holes of several families of MDS codes. In \cite{HYN2021DCC}, Huang \textit{et al.} determined that the dual of a given $(+)$-TGRS code is either another $(+)$-TGRS code or a Han-Zhang code from~\cite{HZ2024}. The Han-Zhang codes form a family of linear codes in which each code is either MDS or NMDS. 
For those which are MDS, it was not investigated whether they are GRS or non-GRS. Starting from the non-GRS MDS properties of $(+)$-TGRS codes, we will soon prove that the Han-Zhang codes are in fact non-GRS in certain cases. Along with the connection established in \cite{WDC2023}, it is natural to investigate the next problem.
\begin{problem}\label{prob.3}
To derive the covering radius and deep holes of Han-Zhang codes in cases where they are non-GRS MDS, can we express $(+)$-ETGRS codes as the second extended codes of $(+)$-TGRS codes?
\end{problem}

\medskip
\noindent
{\bf Our Contributions}
\smallskip

Motivated by Problems \ref{prob.1}, \ref{prob.2}, and \ref{prob.3}, this paper studies properties and decoding of $(+)$-TGRS and $(+)$-ETGRS codes. Our contributions can be summarized as follows: 

\begin{enumerate}
\item We show in Example~\ref{exam.grs} that $[n,\frac{n}{2}]_q$ $(+)$-TGRS codes are not always non-GRS. We proceed to establish three sufficient conditions for such codes to be non-GRS in Theorems \ref{th.non-GRS TGRS for k=n/2111} and \ref{th.non-GRS TGRS for k=n/2222}. 
\item We calculate the Schur squares of duals of $(+)$-ETGRS codes in Theorem \ref{th.Schur square of ETGRS} and use this result to prove that any $\left[n+1,\frac{n+1}{2}\right]_q$ $(+)$-ETGRS code is non-GRS in Theorem \ref{th.non-GRS ETGRS}. Our findings partially address Problem \ref{prob.1} Part 1 and completely resolve Problem \ref{prob.1} Part 2. Examples \ref{exam.non_grs1} and \ref{exam.non_grs_ETGRS} present non-GRS MDS $(+)$-TGRS and $(+)$-ETGRS codes. Applying the theoretical results gives us two classes of non-GRS MDS Hermitian self-dual $(+)$-TGRS codes in Corollaries \ref{coro.HSD_TGRS111} and \ref{coro.HSD_TGRS222}. We prove the nonexistence of a Galois self-dual $(+)$-ETGRS code in Corollary \ref{cor.no_galois_sd_codes}.

\item Using $(+)$-TGRS codes that are non-GRS MDS, we show in Theorem \ref{th.no_ECP_TGRS} that, if $2+\eta \sum_{i=1}^{n}a_i\neq 0$ and $6\leq n\leq q$, then $[n, \frac{n}{2}, \frac{n}{2}+1]_q$ $(+)$-TGRS codes with $n$ divisible by $4$ do not admit $t$-ECPs. This partially answers Problem \ref{prob.2} Part 1. In Theorems \ref{th.ECP111} and \ref{th.ECP222}, we fully resolve Problem \ref{prob.2} Parts 2 and 3 by explicitly constructing ECPs for the specified cases. We develop a better decoding algorithm for $(+)$-ETGRS codes and present it in Theorem \ref{th.decoding ECP} and Algorithm \ref{alg.decoding}. Comparative results, discussed in Remarks \ref{rem.complexity_comparison111} and \ref{rem.complexity_comparison222}, demonstrate that our algorithm outperforms existing methods in certain scenarios. Example \ref{exam.decoding} gives a concrete decoding process for a particular instance for illustration.

\item Addressing Problem \ref{prob.3}, we establish a new connection between $(+)$-ETGRS and $(+)$-TGRS codes by interpreting the former as a second extended code of the latter in Theorem \ref{th.ETGRS is second kind of extended code}. Combining this connection with the results from \cite{WDC2023}, we derive the covering radius and deep holes of the duals of MDS $(+)$-TGRS codes and, consequently, those of MDS Han-Zhang codes in Theorem \ref{th.covering radius and deep holes} and Table \ref{tab:deep_hole}. The fact that these MDS Han-Zhang codes are non-GRS extends a related result in \cite{JMXZ2024}. 
Example \ref{exam.deep_holes} illustrates the relevant results.
\end{enumerate}

In terms of organization, after this introduction, Section \ref{sec.2} reviews basic notations and useful results. Section \ref{sec.3} explores the non-GRS MDS properties of $(+)$-TGRS and $(+)$-ETGRS codes. Section \ref{sec.decoding} focuses on the decoding of the codes in terms of their ECPs and deep holes. Section \ref{sec.conclusion} contains a summary and concluding remarks. 

\section{Preliminaries}\label{sec.2}
We fix the following notations.
\begin{itemize}
\item For a prime $p$ and an $m \in \mathbb{N}$, we denote by $\F_q$ the {\em finite field} of size $q=p^m$.
    
\item $\F_q^n$ is the {\em $n$-dimensional vector space} over $\F_q$.

\item $\F_q^*$ is the {\em multiplicative group} of $\F_q$  and $\eta\in \F_q^*$ is a generic representation of its element.
    
\item For an $[n,k,d]_q$ code $\C$, its {\em dimension} and {\em minimum distance} are denoted respectively by $\dim(\C)$ and $d(\C)$ or $d$ when $\C$ is clear from the context.         
    
\item Given integers $n, \ell \in \mathbb{N}$, we can define a subset $\SSS:=\{a_1,a_2,\ldots,a_n\}\subseteq \F_q$ to construct a vector ${\bf w}=(w_1,w_2,\ldots,w_n)\in (\F_q^*)^n$ such that 
\begin{equation}\label{eq:wi}
w_i :=\prod_{1\leq j\neq i\leq n}(a_i-a_j)^{-1}~{\rm for}~ 1\leq i\leq n.
\end{equation}
For a vector $\mathbf{v}=(v_1,v_2,\ldots,v_n)\in (\F_q^*)^n$, we can define $\vvv^{\ell} := \left(v_1^{\ell},v_2^{\ell},\ldots,v_n^{\ell}\right)$.

\item For any vector ${\bf x}=(x_1,x_2,\ldots,x_n)\in \F_q^n$, we define 
\[
\supp({\bf x}) :=\{i \, : \, x_i\neq 0\}, \mbox{ and } 
z({\bf x}) :=\{i \, : \, x_i=0\}.
\]
    
\item Given vectors ${\bf x}=(x_1,x_2,\ldots,x_n),~{\bf y}=(y_1,y_1,\ldots,y_n)\in \F_q^n$ and an integer $e$ such that $0\leq e\leq m-1$, the {\em $e$-Galois inner product} of ${\bf x}$ and ${\bf y}$ is 
\[
\langle {\bf x}, {\bf y} \rangle_e := \sum_{i=1}^{n}x_i \, y_i^{p^e}.
\]
The {\em $e$-Galois dual} of a given $[n,k]_q$ code $\C$ is 
\[
\C^{\perp_e} :=\{{\bf y}\in \F_q^n \, : \, \langle {\bf x}, {\bf y} \rangle_e=0 \mbox{ for all } {\bf x}\in \C\}.
\]
We use $\C^{\perp}$ and $\C^{\perp_{\rm H}}$ to denote the {\em Euclidean dual code} and {\em Hermitian dual code} of $\C$. They correspond to the cases $e=0$ and $e=\frac{m}{2}$, for even $m$.

\item We denote by $\mathbf{0}$ and $\mathbf{1}$ the {\em zero} and {\em one} row or column vectors of specified length. 

\item We use $[a,b]$, $(a,b]$, and $[a,b)$ to denote three sets of integers $\{x \, : \, a\leq x\leq b\}$,  $\{x \, : \, a< x \leq b\}$, and $\{x \, : \, a\leq x< b\}$, respectively.
    
\item The $\F_q$-\emph{linear space spanned by all vectors} ${\bf x}_1, {\bf x}_2, \ldots, {\bf x}_{\ell}$ is $\langle {\bf x}_1, {\bf x}_2, \ldots, {\bf x}_{\ell} \rangle$.
    
\item Given $q$, $k$, and $\eta$, we define two $k$-dimensional polynomial spaces 
\begin{align*}
\F_q[x]_{k} &:=\left\{f(x)=\sum_{i=0}^{k-1}g_i \, x^i  \, : \, g_i\in \F_q \mbox{ for all } 1\leq i\leq k-1\right\} \mbox{ and}\\
\mathbb{V}_{q}[x]_{k} &:=\left\{f(x)= \sum_{i=0}^{k-1} f_i \, x^i + \eta \, f_{k-1} \, x^{k}  \, : \, f_i \in \F_q \mbox{ for all } 0\leq i\leq k-1\right\}.
\end{align*}
\end{itemize}

Equipped with the above notations, one can define GRS and extended GRS codes. 

\begin{definition}\label{def.GRS code}
Given $q$, $n$, and $k$ such that $1\leq k\leq n\leq q$, an $[n,k,n-k+1]_{q}$ {\em generalized Reed-Solomon (GRS) code} associated with $\SSS$ and $\mathbf{v}$ is 
\begin{align*}
\GRS_k(\SSS,\mathbf{v}) :=  \left\{(v_1 \, f(a_1), v_2 \, f(a_2),\ldots,v_n \, f(a_n)) \, : \,  f(x)\in \F_q[x]_{k}\right\}. 
\end{align*} 
An $[n+1,k,n-k+2]_{q}$ {\em extended GRS, often called EGRS, code} associated with $\SSS$ and $\mathbf{v}$ 
is  
\begin{align*}
\EGRS_k(\SSS,\mathbf{v},\infty) :=  \left\{(v_1 \, f(a_1), v_2 \, f(a_2),\ldots,v_n \, f(a_n),f_{k-1}) \, : \,  f(x)\in \F_q[x]_{k}\right\}.
\end{align*} 
\end{definition}

We know, \textit{e.g.}, from \cite{JX2017} that $\GRS_k(\SSS,\mathbf{v})$ and $\EGRS_k(\SSS,\mathbf{v},\infty)$, respectively, have generator matrices
\begin{align}\label{eq.GRS.generator matrix}
G_{\GRS_k(\SSS,\mathbf{v})} & =
\begin{pmatrix}[1.3]
v_1 & v_2 &  \ldots & v_n\\
v_1 \, a_1 & v_2 \, a_2 & \ldots & v_n \, a_n\\
\vdots  & \vdots  & \ddots & \vdots\\
v_1 \, a^{k-1}_1  & v_2 \, a^{k-1}_2  & \ldots & v_n \, a^{k-1}_n
\end{pmatrix} \mbox{ and}\\
G_{\EGRS_k(\SSS,\mathbf{v},\infty)} & =
\begin{pmatrix}[1.4]
v_1 & v_2 & \ldots & v_n & 0\\
v_1 \, a_1 & v_2 \, a_2 & \ldots & v_n \, a_n & 0\\
\vdots  & \vdots  & \ddots & \vdots & \vdots\\
v_1 \, a^{k-2}_1  & v_2 \, a^{k-2}_2  & \ldots & v_n \, a^{k-2}_n & 0\\
v_1 \, a^{k-1}_1  & v_2 \, a^{k-1}_2  & \ldots & v_n \, a^{k-1}_n & 1
\end{pmatrix}. 
\end{align}

We recall some results on $(+)$-TGRS and $(+)$-ETGRS codes. 

\begin{lemma}{\rm (\!\! \cite[Lemma 2.6]{HYN2021DCC}, \cite[Theorem 3.2]{ZL2024})}\label{lem.parameters_of_ETGRS} 
Let $\mathbf{S}_k =\left\{\sum_{a_i\in I}a_i  \, : \, \mbox{ for all } I\subseteq \SSS \mbox{ such that } |I|=k \right\}$. 
The following statements hold. 
\begin{enumerate}[1.]
\item The code $\TGRS_k(\SSS,\mathbf{v},\eta)$ is an $[n,k,n-k+1]_{q}$ MDS code if and only if $-\eta^{-1} \notin \mathbf{S}_k$.\\
The code $\ETGRS_k(\SSS,\mathbf{v},\eta,\infty)$ is an $[n+1,k,n-k+2]_{q}$ MDS code if and only if $-\eta^{-1} \notin \mathbf{S}_k$.

\item The code $\TGRS_k(\SSS,\mathbf{v},\eta)$ is an 
$[n,k,n-k]_{q}$ NMDS code if and only if $-\eta^{-1} \in \mathbf{S}_k$.\\
The code $\ETGRS_k(\SSS,\mathbf{v},\eta,\infty)$ is an 
$[n+1,k,n-k+1]_{q}$ NMDS code if and only if $-\eta^{-1} \in \mathbf{S}_k$.
\end{enumerate}
\end{lemma}

\begin{lemma}{\rm (\!\! \cite[Remark 2.1 and Theorem 3.1]{ZL2024})} Given $q$, $n$, $k$, $\SSS$, $\mathbf{v}$, and $\eta$, the code $\ETGRS_k(\SSS,\mathbf{v},\eta,\infty)$ has a generator matrix    
\begin{equation}\label{eq.ETGRS.generator matrix}
G_{\ETGRS_k(\SSS,\mathbf{v},\eta,\infty)} = 
\begin{pmatrix}[1.4]
v_1 & v_2 &  \ldots & v_n & 0\\
v_1 \, a_1 & v_2 \, a_2 &  \ldots & v_n \, a_n & 0 \\
\vdots & \vdots &  \ddots & \vdots & \vdots\\
v_1 \, a^{k-2}_1 & v_2 \, a^{k-2}_2 &  \ldots & v_n \, a^{k-2}_n & 0\\
v_1 \, \left(a^{k-1}_1+\eta \, a_1^k\right) & v_2 \, \left(a^{k-1}_2+ \eta \, a_2^k \right) & \ldots & v_n \, \left(a^{k-1}_n + \eta \, a_n^k \right) & 1
\end{pmatrix}
\end{equation} 
and a parity check matrix
\begin{equation}\label{eq.ETGRS.parity-check matrix}
H_{\ETGRS_k(\SSS,\mathbf{v},\eta,\infty)} = 
\begin{pmatrix}[1.4]
\frac{w_1}{v_1} & \frac{w_2}{v_2} &  \ldots & \frac{w_n}{v_n} & 0 \\
\frac{w_1}{v_1} \, a_1 & \frac{w_2}{v_2} \, a_2 &  \ldots & \frac{w_n}{v_n} \, a_n & 0\\
\vdots & \vdots &  \ddots & \vdots & \vdots\\
\frac{w_1}{v_1} \, a^{n-k-2}_1 & \frac{w_2}{v_2} \, a^{n-k-2}_2 &  \ldots & \frac{w_n}{v_n} \, a^{n-k-2}_n & 0 \\
\frac{w_1}{v_1} \, a^{n-k-1}_1 & \frac{w_2}{v_2} \, a^{n-k-1}_2 &  \ldots & \frac{w_n}{v_n} \, a^{n-k-1}_n  & -\eta\\
\frac{w_1}{v_1} \, a^{n-k}_1 & \frac{w_2}{v_2} \, a^{n-k}_2 & \ldots & \frac{w_n}{v_n} \, a^{n-k}_n & -1-\eta \, \sum_{i=1}^{n} a_i
\end{pmatrix}.
\end{equation}
\end{lemma}

The matrix $G_{\TGRS_k(\SSS,\mathbf{v},\eta)}$ obtained by deleting the last column of $G_{\ETGRS_k(\SSS,\mathbf{v},\eta,\infty)}$ in \eqref{eq.ETGRS.generator matrix} generates $\TGRS_k(\SSS,\mathbf{v},\eta)$. More formally, we write  
\begin{align}\label{eq.first kind of extended codes}
G_{\ETGRS_k(\SSS,\mathbf{v},\eta,\infty)}
=\begin{pmatrix}
G_{\TGRS_k(\SSS,\mathbf{v},\eta)} & {\bf g}^{\top}
\end{pmatrix}
\mbox{ with } {\bf g}^{\top}=(0,0,\ldots,0,1)^{\top}.
\end{align}
Conversely, $\ETGRS_k(\SSS,\mathbf{v},\eta,\infty)$ is the code generated by the matrix that is constructed by appending the column vector ${\bf g}^{\top}$ to the above generator matrix of $\TGRS_k(\SSS,\mathbf{v},\eta)$. 

The $(+)$-ETGRS code constructed as above is called a {\em first kind of extended code} in, \textit{e.g.}, \cite{WDC2023} and \cite{WHLD2024}. We prefer to use the term {\em first extended code}. Sun, Ding, and Chen introduced another extension technique by examining each codeword in a given linear code. They named the resulting code the {\em second kind of extended code} in \cite{SDC2024FFA} and \cite{SDC2024DM}. We use a shorter term {\em second extended code} here. 

\begin{definition}{\rm (\!\! \cite{SDC2024FFA,SDC2024DM})}\label{def.extended code 2}
Given an $[n,k,d]_q$ code $\C$ and a vector $\uuu=(u_1,u_2,\ldots,u_n)\in \F_q^n$, the $[n+1,k,\overline{d}]_q$ code 
\begin{equation}\label{eq:second_ext}
\overline{\C}(\uuu) : =\left\{(c_1,c_2,\ldots,c_n,c_{n+1}) \, : \, (c_1,c_2,\ldots,c_n)\in \C \mbox{ and } c_{n+1}=\sum_{i=1}^{n}u_i \, c_i\right\}
\end{equation}
is a {\em second extended code} of $\C$ with $\overline{d}=d$ or $\overline{d}=d+1$. If $\C$ is generated by a matrix $G$, then $\overline{\C}(\uuu)$ is generated by 
\[
\overline{G} = 
\begin{pmatrix} 
G & G \, \uuu^{\top}
\end{pmatrix}.
\]
\end{definition}

\section{MDS Twisted and Extended Twisted GRS Codes are Non-GRS}\label{sec.3}

In this section, we study the non-GRS MDS properties of $(+)$-TGRS and $(+)$-ETGRS codes. We begin by recalling a definition and several known results.  

\begin{definition}
The \emph{Schur product of two linear codes} $\C_1$ and $\C_2$ of the same length is obtained by multiplying their corresponding codewords element-wise and is given by 
\begin{equation}\label{eq:Schur}
\C_1 \star \C_2=\{ {\bf c}_1\star {\bf c}_2 \, : \, {\bf c}_1 \in \C_1 \mbox{ and } {\bf c}_2\in \C_2 \}.
\end{equation}
If $\C_1=\langle \ccc_{1,1}, \ccc_{1,2}, \ldots, \ccc_{1,k}\rangle$ and $\C_2=\langle \ccc_{2,1}, \ccc_{2,2}, \ldots, \ccc_{2,\ell}\rangle$ 
are two $q$-ary linear codes with the same length, then their Schur product $\C_1\star \C_2$ is the $q$-ary  linear code given by 
\begin{align*}
\C_1 \star \C_2 = \langle {\bf c}_{1,i}\star {\bf c}_{2,j} \, : \, 1 \leq i\leq k \mbox{ and } 1\leq j\leq \ell\rangle. 
\end{align*}
If $\C_1=\C_2=\C$, we abbreviate $\C_1\star \C_2$ as $\C^2$ and call it the \emph{Schur square} of $\C$.
\end{definition}

We emphasize that, if $\C$ and $\D$ are equivalent, then $\C^2$ is equivalent to $\D^2$. Thus, one can decide that a linear code is \emph{not} equivalent to a GRS code 
based on the contrapositive of either of the following two lemmas with respect to certain ranges of code dimensions.

\begin{lemma}{\rm (\!\!\cite[Proposition 10]{MMP2013})}\label{lem.GRS square dimension}
If $\C$ is an $[n,k]_q$ GRS code with $3\leq k< \frac{n+1}{2}$, then $\dim(\C^2)= 2k-1.$ 
\end{lemma}

\begin{lemma}{\rm (\!\!\cite[Lemma 3.3]{HL2024DCC})}\label{lem.GRS square distance}
If $\C$ is an $[n,k]_q$ GRS code with $3\leq k< \frac{n+1}{2}$, then $d(\C^2)\geq 2.$
\end{lemma}

\subsection{Non-GRS MDS properties of TGRS codes}
It is known, \textit{e.g.}, from \cite{JX2017}, that the Euclidean dual of a GRS code is also GRS. To prove that an $[n,k]_q$ linear code $\C$ is always non-GRS whenever $3\leq k\leq n-3$, it suffices, based on Lemmas \ref{lem.GRS square dimension} and \ref{lem.GRS square distance}, to show that 
\begin{align}\label{eq:suffice}
\dim(\C^2) &\geq 2k \mbox{ for any } 3\leq k< \frac{n+1}{2} \mbox{ and } \notag\\
d((\C^{\perp})^2) &< 2 \mbox{ for any } \frac{n}{2}\leq k\leq n-3. 
\end{align}
He and Liao in \cite[Lemma 2.4 (c.7)]{HL2023DM} established the sufficiency conditions for $\frac{n}{2}< k\leq n-3$. Beelen \textit{et al.} had earlier confirmed the case for $3\leq k< \frac{n}{2}$ in \cite[Corollary 3]{BPR2022IT}. We address the remaining gap for when $n=2k$. The following example highlights conditions that a $(+)$-TGRS code with parameters $[n, \frac{n}{2}]_q$ must meet to make it non-GRS.
\begin{example}\label{exam.grs}
Using $\SSS=\{0, 3, 4, 5, 9, 10\}\subseteq \F_{11}$, $\vvv={\bf 1}\in (\F_{11}^*)^6$, and $\eta=1$ we generate $\TGRS_3(\SSS,{\bf 1},1)$ by the matrix  
\begin{equation*}
G_{\TGRS_3(\SSS,{\bf 1},1)}=\begin{pmatrix}    
1 & 1 & 1 & 1 & 1 & 1 \\
0 & 3 & 4 & 5 & 9 & 10 \\ 
0 & 3 & 3 & 7 & 7 & 0
\end{pmatrix}. 
\end{equation*}
By Lemma \ref{lem.parameters_of_ETGRS} Part 1, $\TGRS_3(\SSS,{\bf 1},1)$ is a $[6,3,4]_{11}$ MDS code since 
\[
10=-\eta^{-1}\notin \left\{\sum_{a_i\in I}a_i  \, : \, \mbox{ for all } I\subseteq \SSS \mbox{ such that } |I|=3\right\}.
\]
We generate all $[6,3,4]_{11}$ GRS codes by \texttt{MAGMA} \cite{magma} and use its internal $\texttt{IsEquivalent}$ function to check if $\TGRS_3(\SSS,{\bf 1},1)$ is equivalent to any of them. The outputs show that $\TGRS_3(\SSS,{\bf 1},1)$ is equivalent to the $[6,3,4]_{11}$ GRS codes 
\[
\GRS_3(\SSS_1,\vvv),\quad \GRS_3(\SSS_2,\vvv), \quad \GRS_3(\SSS_3,\vvv), \quad \GRS_3(\SSS_4,\vvv)\mbox{, and } \GRS_3(\SSS_5,\vvv)
\]
for any $\vvv\in (\F_{11}^*)^6$, with 
\[
\SSS_1=\{1, 2, 5, 6, 9, 10\}, \quad 
\SSS_2=\{2, 3, 4, 7, 8, 9\}, \quad 
\SSS_3=\{1, 3, 5, 6, 8, 10\}, \quad 
\SSS_4=\{3, 4, 5, 6, 7, 8\}\mbox{, and }
\SSS_5=\{1, 2, 4, 7, 9, 10\}.
\]
\end{example}

\begin{theorem}\label{th.non-GRS TGRS for k=n/2111}
For $k\in [3, \frac{n}{2})\cup (\frac{n}{2},n-3]$, any $(+)$-TGRS code with parameters $[n,k]_q$ is non-GRS. If $2 + \eta \, \sum_{i=1}^{n}a_i \neq 0$ and $6\leq n\leq q$, then any $(+)$-TGRS code with parameters $[n,\frac{n}{2}]_q$ is non-GRS. 
\end{theorem}
\begin{IEEEproof}
We prove that $\TGRS_{\frac{n}{2}}(\SSS,\mathbf{v})$ is non-GRS whenever $n$ is even and $6\leq n\leq q$.  
Using \eqref{eq.ETGRS.generator matrix} and \eqref{eq.first kind of extended codes} we compute  
\begin{align*}
\left(\TGRS_{\tfrac{n}{2}}(\SSS,\mathbf{v})\right)^2 & = 
\left\langle \vvv^2 \star \aaa^{i+j}, \vvv^2 \star 
\left(\aaa^{i+\tfrac{n}{2}-1} + \eta \, \aaa^{i+\tfrac{n}{2}}\right), \vvv^2 \star \left(\aaa^{n-2} + 2 \, \eta \, \aaa^{n-1} + \eta^2 \, \aaa^n \right) \, : \, 0\leq i,j\leq \frac{n}{2}-2 \right\rangle,\\
& = \left \langle \vvv^2 \star \aaa^{r}, \, \vvv^2 \star 
\left(\aaa^{s} + \eta \, \aaa^{s+1}\right), \, \vvv^2\star \left(\aaa^{n-2}+ 2 \, \eta \, \aaa^{n-1} + \eta^2 \, \aaa^n\right) \, : \, 
0\leq r\leq n-4 \mbox{ and } n-4\leq s\leq n-3 \right\rangle,\\
& = \left\langle \vvv^2\star \, \aaa^{r}, \, \vvv^2 \star \left(2 \, \eta \, \aaa^{n-1} + \eta^2 \, \aaa^{n}\right) \, : \, 0\leq r \leq n-2 \right\rangle. 
\end{align*}
Since $0\leq r\leq n-2$, to determine the dimension of $(\TGRS_{\frac{n}{2}}(\SSS,\mathbf{v}))^2$, by \cite[Lemma 9]{BPR2022IT} we need to determine the coefficient of $x^{n-1}$ in the polynomial 
\[
2 \, \eta \, x^{n-1}+ \eta^2 \, x^n - \eta^2 \prod_{i=1}^{n}(x-a_i).
\]
Since both $\eta$ and $2+\eta\sum_{i=1}^{n}a_i$ are nonzero, we have
\[
2 \, \eta + \eta^2 \, \sum_{i=1}^{n}a_i=\eta \left(2+\eta \, \sum_{i=1}^{n}a_i \right)\neq 0.
\]
Hence, $\dim\left(\left(\TGRS_{\frac{n}{2}}(\SSS,\mathbf{v})\right)^2\right)=n>n-1$. By Lemma \ref{lem.GRS square dimension}, we conclude that $\TGRS_{\frac{n}{2}}(\SSS,\mathbf{v})$ is non-GRS. 
\end{IEEEproof}

~\\

We can safely remove the condition $2+ \eta \, \sum_{i=1}^{n}a_i\neq 0$ when $\SSS=\F_q$ or $\F_q^*$ as follows.

\begin{theorem}\label{th.non-GRS TGRS for k=n/2222}
The following statements hold. 
\begin{enumerate}[1.]
    \item For every $k$ and $q$ such that $3\leq k\leq q-3$, any $(+)$-TGRS code with parameters $[q,k]_q$ is non-GRS. 
    \item For every $k$ and $q$ such that $3\leq k\leq q-4$, any $(+)$-TGRS code with parameters $[q-1,k]_q$ is non-GRS. 
\end{enumerate}

\end{theorem}

\begin{IEEEproof}
We first prove that $\TGRS_{\frac{q}{2}}(\F_q,\mathbf{v})$ is non-GRS for even $q \geq 8$. Let 
\[
N\left(\frac{q}{2},-\eta^{-1},\F_{q}\right)
\]
denote the number of subsets $\{x_1,x_2,\ldots,x_{\frac{q}{2}}\}$ of $\F_{q}$ such that $x_1+x_2+\ldots+x_{\frac{q}{2}}=-\eta^{-1}$. By Lemma \ref{lem.parameters_of_ETGRS} Part 2, we obtain
\begin{align*}
\TGRS_{\frac{q}{2}}(\F_q,\mathbf{v}) \mbox{ is an NMDS code } & \iff \eta^{-1}\in \left\{\sum_{a_i\in I}a_i\, : \, \mbox{ for all } I\subseteq \F_q \mbox{ that satisfy } |I|=\frac{q}{2}\right\}\\
& \iff N\left(\frac{q}{2},-\eta^{-1},\F_{q}\right)>0.
\end{align*}
Since $q\geq 8$ is even, $2<\frac{q}{2}<q-2$. By \cite[Corollary 2.8]{LW2008}, we get $N\left(\frac{q}{2},-\eta^{-1},\F_{q}\right)>0$. Thus, $\TGRS_{\frac{q}{2}}(\F_q,\mathbf{v})$ is not an MDS code, making it clearly non-GRS. 
This proves the first claim. To prove the second claim, we follow the same line of argument but using \cite[Corollary 2.7]{LW2008} instead of \cite[Corollary 2.8]{LW2008}. 
\end{IEEEproof}

\begin{remark}
As Example \ref{exam.grs} illustrates, an $[n,\frac{n}{2},\frac{n}{2}+1]_q$ $(+)$-TGRS code may be equivalent to a GRS code. 
In this context, Theorems \ref{th.non-GRS TGRS for k=n/2111} and \ref{th.non-GRS TGRS for k=n/2222} establish three sufficient conditions to decide 
if an $[n, \frac{n}{2}]_q$ $(+)$-TGRS code is indeed non-GRS. 
We have thus partially addressed Problem \ref{prob.1} Part 1.
\end{remark}

Our next example shows a $(+)$-TGRS code which is MDS but non-GRS.
\begin{example}\label{exam.non_grs1}
Let $\SSS=\{2, 5, 7, 9, 10, 12\}\subseteq \F_{13}$, $\vvv={\bf 1}\in (\F_{13}^*)^6$, and $\eta=1$. Consider  the $(+)$-TGRS code $\TGRS_3(\SSS,{\bf 1},1)$ with generator matrix  
\begin{equation*}
G_{\TGRS_3(\SSS,{\bf 1},1)}=
\begin{pmatrix}    
1 & 1 & 1 & 1 & 1 & 1 \\
2 & 5 & 7 & 9 & 10 & 12 \\ 
12 & 7 & 2 & 4 & 8 & 0
\end{pmatrix}. 
\end{equation*}
By Lemma \ref{lem.parameters_of_ETGRS} Part 1, since 
\[
12=-\eta^{-1} \notin \left\{\sum_{a_i \in I} a_i  \, : \, \mbox{ for all } I\subseteq \SSS_2 \mbox{ such that } |I|=3\right\},
\]
we know that $\TGRS_3(\SSS,{\bf 1},1)$ is a $[6,3,4]_{13}$ MDS code. We also verify that $2+\eta \, \sum_{i=1}^{6}a_i=8\neq 0$. By generating all $[6,3,4]_{13}$ GRS codes and checking them using the $\texttt{IsEquivalent}$ function in \texttt{MAGMA} \cite{magma}, we confirm that $\TGRS_3(\SSS,{\bf 1},1)$ is not equivalent to 
any $[6,3,4]_{13}$ GRS code as guaranteed by Theorem \ref{th.non-GRS TGRS for k=n/2111}.  
\end{example}

\subsection{Non-GRS MDS properties of ETGRS codes}

Let $n$ and $k$ be such that $3\leq k\leq n-2$. For an extended code $\C$ with parameters $[n+1,k]_q$, Lemmas \ref{lem.GRS square dimension} and \ref{lem.GRS square distance} imply that $\C$ is always non-GRS if one can show that 
\begin{align}\label{eq:suffice222}
\dim(\C^2) &\geq 2k \mbox{ for any } 3\leq k< \frac{n+2}{2} \mbox{ and } \notag\\
d((\C^{\perp})^2) &< 2 \mbox{ for any } \frac{n+1}{2}\leq k\leq n-2. 
\end{align}
Zhu and Liao established these conditions only for $k\in [3,\frac{n}{2}] \cup [\frac{n}{2}+1,n-2]$ in the proof of \cite[Theorem 3.4]{ZL2024}. 
A gap remained for the case where $n$ is odd and $2k=n+1$. Zhu and Liao also determined $\ETGRS_k(\SSS, \mathbf{v}, \eta, \infty)^2$ and 
$(\ETGRS_k(\F_q, \mathbf{v}, \infty)^2)^{\perp}$ in \cite[Lemma 3.2]{ZL2024} and \cite[Lemma 4.1]{ZL2024}, respectively. The Schur square $(\ETGRS_k(\SSS, \mathbf{v}, \eta, \infty)^{\perp})^2$ remained undetermined due to the fact that, in general, 
$(\C^2)^{\perp} \neq (\C^{\perp})^2$ and $\ETGRS_k(\SSS,\mathbf{v},\eta,\infty)^{\perp}$ is not necessarily a $(+)$-ETGRS code.       

We now determine the Schur square of $\ETGRS_k(\SSS, \mathbf{v}, \eta, \infty)^{\perp}$ and use it to assess whether $\ETGRS_k(\SSS, \mathbf{v}, \eta, \infty)$ is non-GRS for all cases.

\begin{theorem}\label{th.Schur square of ETGRS}
For any $\eta\in \F_q^*$ and suitable $n$ and $k$, the following statements hold.
\begin{enumerate}[1.]
\item If $3\leq k\leq \frac{n+1}{2}$, then 
$\displaystyle{
(\ETGRS_k(\SSS,\mathbf{v},\eta,\infty)^{\perp})^2=\F_q^{n+1}}$.
\item  If $\frac{n+1}{2}< k\leq n-2$, then $\displaystyle{
(\ETGRS_k(\SSS,\mathbf{v},\eta,\infty)^{\perp})^2= \C_1+\C_{2n-2k+1}}$,
with 
\[
\C_1=\langle (0,0,\ldots,0,1)\rangle \mbox{ and } 
\C_{2n-2k+1}=\overline{\GRS_{2n-2k+1}\left(\SSS,\frac{\www^2}{\vvv^2}\right)}({\bf 0}).
\]
\end{enumerate}
\end{theorem}

\begin{IEEEproof}
Let $n$, $k$, and $h$ be such that $0\leq h\leq n-k$. Setting 
\[
r_h := 
\begin{cases}
0, & \mbox{if } 0\leq h\leq n-k-2,\\
-\eta, & \mbox{if } h=n-k-1,\\
-1-\eta\sum_{i=1}^{n}a_i, & \mbox{if } h=n-k,
\end{cases}
\]
we define
\begin{align*}
\frac{\www}{\vvv} & := \left(\frac{w_1}{v_1},\frac{w_2}{v_2},\ldots,\frac{w_n}{v_n}\right),\\
\rrr^h &:=\left(\frac{w_1}{v_1} \, a_1^h, \frac{w_2}{v_2} \, a_2^h,\ldots,\frac{w_n}{v_n} \, a_n^h\right) \mbox{, and}\\
(\rrr^h,r_h) & :=\left(\frac{w_1}{v_1} \, a_1^h, \frac{w_2}{v_2} \, a_2^h,\ldots,\frac{w_n}{v_n} \, a_n^h, r_h\right).
\end{align*}

We use \eqref{eq.ETGRS.parity-check matrix} to obtain 
\begin{align*}
\left(\ETGRS_k(\SSS,\mathbf{v},\eta,\infty)^{\perp}\right)^2 
&= \left \langle(\rrr^h,0)\star(\rrr^m,r_m),~(\rrr^{n-k-1},r_{n-k-1})\star(\rrr^{n-k-1},r_{n-k-1}),
(\rrr^{n-k-1},r_{n-k-1})\star (\rrr^{n-k},r_{n-k}), \right.\\
 & \qquad \left. (\rrr^{n-k},r_{n-k})\star (\rrr^{n-k},r_{n-k}) \, : \, 0 \leq h\leq n-k-2 \mbox{ and } 0\leq m\leq n-k \right\rangle \\
&= \Bigg\langle \left(\frac{\www}{\vvv} \star \rrr^s,0 \right),~ \left(\frac{\www}{\vvv}\star \rrr^{2n-2k2} , \eta^2 \right),~ 
\left( \frac{\www}{\vvv}\star \rrr^{2n-2k-1}, \eta \left(1+\eta \, \sum_{i=1}^{n}a_i \right)\right), \Bigg. \\
& \qquad \left. \left(\frac{\www}{\vvv}\star \rrr^{2n-2k},\left(1+\eta \,  \sum_{i=1}^{n}a_i\right)^2\right) \, : \, 0\leq s\leq 2n-2k-2 \right\rangle\\
&=\left\langle\left(\frac{\www}{\vvv}\star \rrr^s,0\right),~(0,0,\ldots,0,1),~\left(\frac{\www}{\vvv}\star \rrr^{2n-2k-1},0\right),~\left(\frac{\www}{\vvv}\star \rrr^{2n-2k},0\right) \, : \,
0\leq s\leq 2n-2k-2 \right\rangle\\
&=\left\langle\left(\frac{\www}{\vvv}\star \rrr^s,0\right),~(0,0,\ldots,0,1) \, : \, 0\leq s\leq 2n-2k \right\rangle \\
&=\C_1+ \left\langle\left(\frac{\www}{\vvv}\star \rrr^s,0\right) \, : \, 0\leq s\leq 2n-2k \right\rangle.
\end{align*}
We consider two cases.

\noindent
{\textbf{Case 1:} When $3\leq k\leq \frac{n+1}{2}$}, we have two facts. The first one is that $(\ETGRS_k(\SSS,\mathbf{v},\eta,\infty)^{\perp})^2\subseteq \F_q^{n+1}$. Since $3\leq k\leq \frac{n+1}{2}$, we get $2n-2k \geq n-1$, leading to the fact that 
\begin{align*}
\F_q^{n+1} & = \left\langle\F_q^n\times \{0\},~(0,0,\ldots,0,1) \right\rangle \\ 
& = \left\langle \left(\frac{\www}{\vvv}\star \rrr^s, 0\right),~(0,0,\ldots,0,1) \, : \,  0\leq s\leq n-1 \right\rangle\\
& \subseteq  \left\langle \left(\frac{\www}{\vvv}\star \rrr^s, 0\right),~(0,0,\ldots,0,1) \, : \, 0\leq s\leq 2n-2k \right\rangle\\
& = \left(\ETGRS_k(\SSS,\mathbf{v},\eta,\infty)^{\perp}\right)^2.
\end{align*}
Combining the two, we get $(\ETGRS_k(\SSS,\mathbf{v},\eta,\infty)^{\perp})^2 = \F_q^{n+1}$, confirming the first assertion.

\noindent
{\textbf{Case 2:} When $\frac{n+1}{2}< k\leq n-2$}, we have $2n-2k\leq n-2<n-1$ and 
\begin{align*}
\left\langle \left(\frac{\www}{\vvv}\star \rrr^s,0\right) \, : \, 0\leq s\leq 2n-2k \right \rangle 
&= \left\{\left(\frac{w_1^2}{v_1^2} \, g(a_1), \frac{w_2^2}{v_2^2} \, g(a_1),\ldots, \frac{w_n^2}{v_n^2} \, g(a_1),0\right) \, : \, g(x)\in \F_q[x]_{2n-2k+1} \right\} \\
&= \overline{\GRS_{2n-2k+1}\left(\SSS,\frac{\www^2}{\vvv^2}\right)}({\bf 0}).
\end{align*}
Hence, $(\ETGRS_k(\SSS,\mathbf{v},\eta,\infty)^{\perp})^2=\C_1+\C_{2n-2k+1},$ which gives us the second claim.
\end{IEEEproof}

We have shown that $\ETGRS_k(\SSS,\mathbf{v},\eta,\infty)$ is non-GRS for any $3\leq k\leq n-2$, which improves on \cite[Theorem 3.4]{ZL2024}. 

\begin{theorem}{\rm (An improvement on \cite[Theorem 3.4]{ZL2024})}\label{th.non-GRS ETGRS}
If $n$ and $k$ are positive integers such that $3\leq k\leq n-2$, then any $[n+1,k]_q$ MDS $(+)$-ETGRS code is non-GRS. 
\end{theorem}
\begin{IEEEproof}
For $3\leq k\leq \frac{n}{2}$, any $\ETGRS_k(\SSS,\mathbf{v},\eta,\infty)$ has been shown to be non-GRS in \cite[Theorem 3.4 ({\bf Case 1})]{ZL2024}. Whenever $\frac{n+1}{2}\leq k\leq n-2$, we have $3\leq n+1-k\leq \frac{n+1}{2}$. By Theorem \ref{th.Schur square of ETGRS}, $\displaystyle{
d((\ETGRS_k(\SSS,\mathbf{v},\eta,\infty)^{\perp})^2)=1<2}$. By Lemma \ref{lem.GRS square distance}, we conclude that $\ETGRS_k(\SSS,\mathbf{v},\eta,\infty)$ is also non-GRS.
\end{IEEEproof}

\begin{remark} The following assertions can be immediately verified based on the results we have presented thus far.
\begin{enumerate}[1.]
\item Theorem \ref{th.non-GRS ETGRS} provides a complete solution to Problem \ref{prob.1} Part 2.
\item By Theorem \ref{th.Schur square of ETGRS} Statement 2, the code $(\ETGRS_k(\SSS,\mathbf{v},\eta,\infty)^{\perp})^2$ is \emph{not} the whole space $\F_q^{n+1}$ for $\frac{n+2}{2}\leq k\leq n-2$. In this case, deleting the last coordinate from each codeword of $(\ETGRS_k(\SSS,\mathbf{v},\eta,\infty)^{\perp})^2$ yields an $[n,2n-2k+1,2k-n]_q$ GRS code $\GRS_{2n-2k+1}\left(\SSS,\frac{\www^2}{\vvv^2}\right)$. This procedure may have some interesting applications in private information retrieval schemes \cite{BHMR2025}, linear secret sharing, and secure multi-party computation protocols \cite{CCMPX2015}.
\end{enumerate}
\end{remark}

Our next example shows a $(+)$-ETGRS code which is non-GRS MDS.

\begin{example}\label{exam.non_grs_ETGRS}
Let $\SSS=\{1, 5, 6, 9, 10\}\subseteq \F_{11}$, $\vvv={\bf 1}\in (\F_{11}^*)^5$, and $\eta=3$. The $(+)$-ETGRS code $\ETGRS_3(\SSS,{\bf 1},3,\infty)$ is generated by 
\begin{equation*}
G_{\ETGRS_3(\SSS,{\bf 1},3,\infty)}=
\begin{pmatrix}    
1 & 1 & 1 & 1 & 1 & 0 \\
1 & 5 & 6 & 9 & 10 & 0 \\ 
4 & 4 & 2 & 2 & 9 & 1
\end{pmatrix}. 
\end{equation*}
Since 
\[
7=-\eta^{-1} \notin \left\{\sum_{a_i\in I}a_i  \, : \, \mbox{ for all } I \subseteq \SSS \mbox{ such that } |I|=3\right\},
\]
we invoke Lemma \ref{lem.parameters_of_ETGRS} Part 1 to conclude that $\ETGRS_3(\SSS,{\bf 1},3,\infty)$ is a $[6,3,4]_{11}$ MDS code. We generate all $[6,3,4]_{11}$ GRS codes by \texttt{MAGMA} \cite{magma} and check for equivalence. The outputs show that $\ETGRS_3(\SSS,{\bf 1},3,\infty)$ is not equivalent to any $[6,3,4]_{11}$ GRS code in agreement with Theorem \ref{th.non-GRS ETGRS}.
\end{example}

\subsection{Applications of TGRS and ETGRS Codes} 

Guo \textit{et al.} constructed two families of Hermitian self-dual $(+)$-TGRS codes in \cite[Theorems 5 and 6]{GLLS2023} and analyzed their MDS properties in \cite[Corollaries 7 and 8]{GLLS2023}. \emph{Whether the codes are GRS or not was not investigated.} We combine their results with Theorem \ref{th.non-GRS TGRS for k=n/2111} to construct the following two families of non-GRS MDS Hermitian self-dual $(+)$-TGRS codes. 

\begin{corollary}\label{coro.HSD_TGRS111}
Let $q$, $n$, and $k$ be given such that $n=2k\leq q$. Let $\w\in \F_{q^2}\setminus \F_q$ be a primitive element of $\F_{q^2}$. Let $a,x_i\in \F_q$ for $1\leq i\leq n$ and let 
$\SSS=\{a \, \w + x_1, a \, \w + x_2,\ldots,a \, \w + x_n\}$. Let $\eta\in \F_{q^2}^*$ satisfy $\eta+\eta^q=0$. If either $\sum_{i=1}^{n}(a\w+x_i)=0$ or $q=2^s$ fails to hold and $2+\eta \sum_{i=1}^{n}(a\w+x_i)\neq 0$, then there exists a $\vvv\in (\F_{q^2}^*)^n$ such that $\TGRS_k(\SSS,\mathbf{v},\eta)$ is an $[n,\frac{n}{2},\frac{n}{2}+1]_{q^2}$ non-GRS MDS Hermitian self-dual code.
\end{corollary}
\begin{IEEEproof}
The desired conclusion follows immediately from \cite[Theorem 5 and Corollary 8]{GLLS2023} and the above Theorem \ref{th.non-GRS TGRS for k=n/2111}.
\end{IEEEproof}

\begin{corollary}\label{coro.HSD_TGRS222}
Let $q$, $n$, and $k$ be given such that $n=2k\leq q$. Let $\w\in \F_{q^2}\setminus \F_q$ be a primitive element of $\F_{q^2}$. Let $a,x_i\in \F_q$ for $1\leq i\leq n$. Let $m$ be such that with $1\leq m\leq q$ and let $\SSS = \{a+ \w^m \, x_1, a+ \w^m \, x_2, \ldots, a+ \w^m \, x_n\}$. Let $\eta\in \F_{q^2}^*$ satisfy $\mu \, \eta-\eta^q=0$ for some $\mu\in \F_{q^2}$. 
If either $\sum_{i=1}^{n} (a+\w^m \, x_i)=0$ or $q=2^s$ fails to hold and $2+\eta \, \sum_{i=1}^{n}(a+\w^m \, x_i) \neq 0$, then there exists a $\vvv\in (\F_{q^2}^*)^n$ such that $\TGRS_k(\SSS,\mathbf{v},\eta)$ is an $[n,\frac{n}{2},\frac{n}{2}+1]_{q^2}$ non-GRS MDS Hermitian self-dual code.
\end{corollary}

\begin{IEEEproof}
The claim follows by combining \cite[Theorem 6 and Corollary 8]{GLLS2023} and our Theorem \ref{th.non-GRS TGRS for k=n/2111}.
\end{IEEEproof}

As Guo \textit{et al.} mentioned in \cite[Remark 3]{GLLS2023}, Huang \textit{et al.} have previously proved in \cite[Theorem 2.8]{HYN2021DCC} that there is no Euclidean self-dual $(+)$-TGRS code if $\sum_{i=0}^{n}a_i=0$ but there is a Hermitian self-dual $(+)$-TGRS code. Additionally, for $q$-ary Gabidulin codes with even $q$, Nebe and Willems proved the nonexistence of \emph{Euclidean} self-dual Gabidulin codes in \cite{NW2016}. More recently, Islam and Horlemann in \cite{IH2023} considered self-duality of such codes under $e$-Galois inner products. Their investigation revealed the existence of \emph{Hermitian} self-dual Gabidulin codes. Zhu and Liao showed in \cite[Corollary 4.4]{ZL2024} that no Euclidean self-dual $(+)$-ETGRS code exists. 

We use the fact that $(+)$-TGRS MDS codes are non-GRS to decide on the nonexistence of $e$-Galois self-dual $(+)$-ETGRS codes.

\begin{corollary}\label{cor.no_galois_sd_codes}
There does not exists any $e$-Galois self-dual 
$\left[n+1,\frac{n+1}{2}\right]_q$ $(+)$-ETGRS code.
\end{corollary}

\begin{IEEEproof}
We know from \cite[Theorem 6]{ZW2025} that for a $\ETGRS_{\frac{n+1}{2}}(\SSS,\mathbf{v},\eta,\infty)$ to be an $\left[n+1,\frac{n+1}{2}\right]_q$ $e$-Galois self-dual code, 
$\TGRS_{\frac{n+1}{2}}(\SSS,\mathbf{v},\eta)$ must be an 
$[n,\frac{n+1}{2}]_q$ GRS code. Since $\frac{n+1}{2}>\frac{n}{2}$, our claim follows immediately from Theorem \ref{th.non-GRS TGRS for k=n/2111}.
\end{IEEEproof}

\begin{remark}\label{rem.non-GRS ETGRS} Some remarks are in order regarding the results discussed in this subsection. 
\begin{enumerate}[1.]
\item Huang \textit{et al.} constructed four families of Euclidean self-dual $(+)$-TGRS codes in \cite{HYN2021DCC}. 
Just like in Corollaries \ref{coro.HSD_TGRS111} and \ref{coro.HSD_TGRS222}, we can obtain four families of 
non-GRS MDS Euclidean self-dual $(+)$-TGRS codes by combining Theorem \ref{th.non-GRS TGRS for k=n/2111} with \cite{HYN2021DCC}. 
We refrain from supplying the details for brevity.

\item Zhu and Wan fixed a slightly different notation than ours in \eqref{eq:space}. In defining a TGRS code, they used
\[
\mathbb{V}_{k,t,h,\eta,q}:= \left\{ f(x) = \sum_{i=0}^{k-1} f_i \, x^i + \eta \, f_h \, x^{k-1+t} : f_i \in \mathbb{F}_q \mbox{ for all } 0 \leq i \leq k-1 \right\}
\]
to construct, for a given $h$,
\[
{\rm TGRS}_{k,t}(\mathcal{S},\mathbf{v}, \eta) = 
\left\{(v_1f(a_1),v_2f(a_2),\ldots,v_nf(a_n)) \, : \,  f(x)\in \mathbb{V}_{k,t,h,\eta,q} \right\}.
\]
They used the fact that $\TGRS_{\frac{n+1}{2},t}(\SSS,\mathbf{v},\eta)$ is non-GRS to prove the nonexistence of $e$-Galois self-dual ETGRS codes for odd $n \geq 11$ and $3\leq t\leq \frac{n-5}{2}$, with $h=\frac{n-1}{2}$, in \cite[Theorem 7]{ZW2025}. When $t=1$, their $\TGRS_{\frac{n+1}{2},1}(\SSS,\mathbf{v},\eta)$ coincides with our $\TGRS_{\frac{n+1}{2}}(\SSS,\mathbf{v},\eta)$. It is, therefore, clear that the result in Corollary \ref{cor.no_galois_sd_codes} does not overlap with the known one from~\cite[Theorem 7]{ZW2025}.
\end{enumerate}
\end{remark}

\section{Faster Decoding}\label{sec.decoding}

This section addresses Problems \ref{prob.2} and \ref{prob.3}. Given a $(+)$-TGRS code and its extension, we determine their $t$-ECPs explicitly once existence can be confirmed. Switching focus to non-GRS MDS Han-Zhang codes, we infer their covering radii and describe a family of deep holes for the codes based on the duals of $(+)$-TGRS codes.

\subsection{Decoding TGRS codes by their ECPs}

This subsection focuses on Problem \ref{prob.2} Part 1. 
A $t$-ECP of any linear code $\C$ is a pair of linear codes $(\mathcal{A},\mathcal{B})$ that is useful to design a general decoding framework that corrects up to $t$ errors in polynomial time. We know that $t\leq \left \lfloor \frac{d(\C)-1}{2} \right \rfloor$ from \cite{P1992}. The following is an alternative definition of the $t$-ECP.
 
\begin{definition}{\rm (\!\! \cite{DK1994,P1992})}\label{def.ECP}
Let $\C$ be an $[n,k]_q$ linear code. Let $\mathcal{A}$ and $\mathcal{B}$ be two linear codes with the same length $n$ over $\F_{q^\ell}$ for a given $\ell \in \mathbb{N}$. If \emph{all} of the following conditions hold, then $(\mathcal{A},\mathcal{B})$ is a {\em $t$-error-correcting pair} or {\em $t$-ECP} of $\C$. 
\begin{enumerate}
\item $\mathcal{A} \star \mathcal{B}\subseteq \C^{\perp}$.
\item $d(\mathcal{B}^{\perp})> t$.
\item $\dim(\mathcal{A})> t$.
\item $d(\mathcal{A})+d(\C)> n$.
\end{enumerate}
\end{definition}

\begin{lemma}{\rm (\!\! \cite[Theorem 6.2]{lem_ECP})}\label{lem.GRS_ECP}
Let $\C$ be an $[n,n-2t,2t+1]_q$ MDS code for an integer $t \in \left[1,\frac{n}{2}-1\right]$. The code $\C$ has a $t$-ECP $(\mathcal{A},\mathcal{B})$ over a finite extension field of $\F_{q}$ if and only if $\C$ is a GRS code.  
\end{lemma}

\begin{theorem}\label{th.no_ECP_TGRS}
Let $n \in [8,q]$ be divisible by $4$, written as $n=4 \, \ell$. If $2+ \eta \, \sum_{i=1}^{n} a_i \neq 0$, then any $(+)$-TGRS code with parameters $\left[n,\frac{n}{2},\frac{n}{2}+1\right]_q$ has no $\ell$-ECP.
\end{theorem}

\begin{IEEEproof}
Let $\C$ be a $(+)$-TGRS code and let $t=\left\lfloor \frac{d(\C)-1}{2} \right\rfloor$. Writing $n=4 \, \ell$, with $2\leq \ell\leq \frac{n}{2}-1$, the code $\C$ has parameters $[4 \, \ell, 2 \, \ell, 2 \, \ell+1]_q$ and $2 \leq t=\left\lfloor \frac{2\ell}{2} \right\rfloor = \frac{n}{4}\leq \frac{n}{2}-1$. For a contradiction, let us assume that $\C$ has an $\ell$-ECP. By Lemma \ref{lem.GRS_ECP}, $\C$ must be a GRS code. This contradicts the non-GRS property of 
$\C$ shown in Theorem \ref{th.non-GRS TGRS for k=n/2111}. Thus, under the stated conditions, an $[n,\frac{n}{2},\frac{n}{2}+1]_q$ $(+)$-TGRS code cannot have $\ell$-ECP. 
\end{IEEEproof}

\begin{remark}
Theorem \ref{th.no_ECP_TGRS} gives a partial solution to Problem \ref{prob.2} Part 1.
\end{remark}

\subsection{Decoding ETGRS codes by their ECPs}
As mentioned earlier, Li \textit{et al.} proved that, when $n-k$ is odd, no $[n+1,k,n-k+2]_q$ $(+)$-ETGRS code has $t$-ECPs, with 
\begin{equation}\label{eq:t}
t=\left \lfloor \frac{d(\ETGRS_{k}(\SSS,{\bf v},\eta,\infty))-1}{2} \right \rfloor.
\end{equation}
They also discussed the existence of $t$-ECPs of some MDS codes of lengths strictly less than the field size in \cite[Theorem 11]{LZS2025}. Prior to our present work, \emph{the existence and explicit forms of $t$-ECPs remained unknown for general $(+)$-ETGRS codes.} To address Problem \ref{prob.2} Part 2, we show that any $(+)$-ETGRS code with parameters $[n+1, k, n - k + 2]_q$ has a $(t - 1)$-ECP when $n - k$ is odd and a $t$-ECP when $n - k$ is even. 
We also prove that any $(+)$-ETGRS code with parameters $[n+1, k, n - k + 1]_q$ has a $t$-ECP. 
We now establish the existance of ECPs for odd $n-k$ and construct them explicitly.

\begin{theorem}\label{th.ECP111}
If $n-k$ is odd, then any $(+)$-ETGRS code $\ETGRS_k(\SSS,\mathbf{v},\eta,\infty)$ has an $\frac{n-k-1}{2}$-ECP $(\mathcal{A},\mathcal{B})$ with 
\[
\mathcal{A}=\overline{\GRS_{\frac{n-k+1}{2}}(\SSS,{\bf 1})}({\bf 0}) \mbox{ and } \mathcal{B}=\EGRS_{\frac{n-k-1}{2}}\left(\SSS,\frac{\www}{\vvv},\infty\right).
\]
\end{theorem}
\begin{IEEEproof}
Let $t$ be as given in \eqref{eq:t}. Since $n-k$ is odd, we infer by Lemma \ref{lem.parameters_of_ETGRS} that 
\[
\frac{n-k-1}{2}= 
\begin{cases}
\left \lfloor \frac{n-k+1}{2} \right \rfloor-1=t-1 \mbox{, if } \eta \notin \left\{\sum_{a_i\in I}a_i  \, : \, \mbox{ for all } I\subseteq \SSS \mbox{ such that } |I| = k \right\},\\
\left \lfloor \frac{n-k}{2} \right \rfloor=t \mbox{, if } \eta \in \left\{\sum_{a_i \in I}a_i  \, : \, \mbox{ for all } I\subseteq \SSS \mbox{ such that } |I|=k\right\}.
\end{cases}
\]
We note that $1\leq \frac{n-k-1}{2}< \frac{n+1}{2}-1$ by taking into account the conditions that $3\leq k\leq n-2$ and $n-k$ is odd.
Our task is to show that any $(+)$-ETGRS code $\ETGRS_k(\SSS,\mathbf{v},\eta,\infty)$ has an $\frac{n-k-1}{2}$-ECP when $n-k$ is odd. 

From \eqref{eq.GRS.generator matrix}, we infer that $\mathcal{A}= \overline{\GRS_{\frac{n-k+1}{2}}(\SSS,{\bf 1})}({\bf 0})$ and $\mathcal{B}= \EGRS_{\frac{n-k-1}{2}}(\SSS,\frac{\www}{\vvv},\infty)$ have respective generator matrices 
\begin{align}\label{eq.ECP_GM111}
G_{\mathcal{A}}=\begin{pmatrix}[1.4]  
1 &  1 &   \ldots & 1 & 0 \\ 
a_1  & a_2  &  \ldots & a_{n} & 0 \\
\vdots  & \vdots  &  \ddots & \vdots & \vdots \\
a^{\frac{n-k-1}{2}}_1  & a^{\frac{n-k-1}{2}}_2  &  \ldots & a^{\frac{n-k-1}{2}}_{n} & 0 
\end{pmatrix} \mbox{ and }
G_{\mathcal{B}}=\begin{pmatrix}[1.4]
\frac{w_1}{v_1}  & \frac{w_2}{v_2}  &  \ldots & \frac{w_n}{v_n} & 0 \\ 
\frac{w_1}{v_1} \, a_1    & \frac{w_2}{v_2} \, a_2    &  \ldots & \frac{w_n}{v_n} \, a_{n} & 0 \\
\vdots & \vdots &   \ddots & \vdots & \vdots \\
\frac{w_1}{v_1} \, a_1^{\frac{n-k-5}{2}}    & \frac{w_2}{v_2} \, a_2^{\frac{n-k-5}{2}}    & \ldots & \frac{w_n}{v_n} \, a_{n}^{\frac{n-k-5}{2}} & 0 \\
\frac{w_1}{v_1} \, a_1^{\frac{n-k-3}{2}}    &  \frac{w_2}{v_2} \, a_2^{\frac{n-k-3}{2}}   & \ldots & \frac{w_n}{v_n} \, a_{n}^{\frac{n-k-3}{2}} & 1 
\end{pmatrix}.
\end{align}
Hence, $\mathcal{A}\star \mathcal{B}$ is an $[n+1,n-k-1]_q$ linear code with a generator matrix
\[
G_{\mathcal{A}\star \mathcal{B}}=
\begin{pmatrix}[1.5]
\frac{w_1}{v_1}  &  \frac{w_2}{v_2}  &  \ldots & \frac{w_n}{v_n} & 0 \\ 
\frac{w_1}{v_1} \, a_1    & \frac{w_2}{v_2} \, a_2    & \ldots & \frac{w_n}{v_n} \, a_{n} & 0 \\
\vdots &  \vdots &   \ddots & \vdots & \vdots \\
\frac{w_1}{v_1} \, a_1^{n-k-3}    & \frac{w_2}{v_2} \, a_2^{n-k-3}    & \ldots & \frac{w_n}{v_n} \, a_{n}^{n-k-3} & 0 \\
\frac{w_1}{v_1} \, a_1^{n-k-2}    & \frac{w_2}{v_2} \, a_2^{n-k-2}    & \ldots & \frac{w_n}{v_n} \, a_{n}^{n-k-2} & 0             
\end{pmatrix}.
\]
Since a parity-check matrix of a linear code is a generator matrix of its Euclidean dual, by (\ref{eq.ETGRS.parity-check matrix}) we arrive at 
$\mathcal{A}\star \mathcal{B}\subseteq \ETGRS_{k}(\SSS,{\bf v},\eta,\infty)^{\perp}$, 
implying that the first condition in Definition \ref{def.ECP} holds. 

We confirm that 
\begin{align*}
d(\mathcal{B}^{\perp})&=\frac{n-k-1}{2}+1>\frac{n-k-1}{2}, \quad \dim(\mathcal{A})=\frac{n-k+1}{2}>\frac{n-k-1}{2} \mbox{ and}\\ 
d(\mathcal{A})+d(\ETGRS_{k}(\SSS,{\bf v},\eta,\infty)) 
&= n-\frac{n-k+1}{2}+1+ 
\begin{cases}
n-k+2 \mbox{, if } \eta \notin \left\{\sum_{a_i\in I} \, a_i  \, : \, \mbox{ for all } I \subseteq \SSS \mbox{ such that } |I|=k \right\}\\
n-k+1 \mbox{, if } \eta \in \left\{\sum_{a_i\in I} \, a_i  \, : \, \mbox{ for all } I\subseteq \SSS \mbox{ such that } |I|=k \right\}
\end{cases} \\
& > n+1.
\end{align*}
Hence, the remaining three conditions in Definition \ref{def.ECP} also hold. Based on Definition \ref{def.ECP}, we conclude that 
\[
(\mathcal{A},\mathcal{B})=
\left(\overline{\GRS_{\frac{n-k+1}{2}}(\SSS,{\bf 1})}({\bf 0}),
\EGRS_{\frac{n-k-1}{2}}\left(\SSS,\frac{\www}{\vvv},\infty\right)
\right)
\]
is an $\frac{n-k-1}{2}$-ECP of $\ETGRS_{k}(\SSS,{\bf v},\eta,\infty)$.
\end{IEEEproof}

\begin{theorem}\label{th.ECP222}
If $n-k$ is even and $\eta\in \F_q^*$, then any $(+)$-ETGRS code $\ETGRS_k(\SSS,\mathbf{v},\eta,\infty)$ has an $\frac{n-k}{2}$-ECP $(\mathcal{A},\mathcal{B})$ with 
\begin{equation}\label{eq:AandB}
\mathcal{A}=\EGRS_{\frac{n-k+2}{2}}(\SSS,{\bf 1},\infty)~{\rm and}~
\mathcal{B}=\EGRS_{\frac{n-k}{2}}\left(\SSS,\frac{\www}{\vvv},\infty\right) \, \diag(\underbrace{1,1,\ldots,1}_n,-\eta).
\end{equation}
\end{theorem}

\begin{IEEEproof}
Since $n-k$ is even, by Lemma \ref{lem.parameters_of_ETGRS} we get  
\[
1\leq t=\left \lfloor \frac{d(\ETGRS_{k}(\SSS,{\bf v},\eta,\infty))-1}{2} \right \rfloor=\frac{n-k}{2}<\frac{n+1}{2}-1
\]
for any $(+)$-ETGRS code $\ETGRS_k(\SSS,\mathbf{v},\eta,\infty)$. 
Taking the $\mathcal{A}$ and $\mathcal{B}$ specified in \eqref{eq:AandB} and using \eqref{eq.GRS.generator matrix}, we have the respective generator matrices  
\begin{align}\label{eq.ECP_GM222}
G_{\mathcal{A}}=\begin{pmatrix}[1.3]
1 &  1 &   \ldots & 1 & 0 \\ 
a_1  &  a_2  &  \ldots & a_{n} & 0 \\
\vdots  & \vdots  &  \ddots & \vdots & \vdots \\
a^{\frac{n-k}{2}}_1  & a^{\frac{n-k}{2}}_2  &  \ldots & a^{\frac{n-k}{2}}_{n} & 1
\end{pmatrix} \mbox{ and }
G_{\mathcal{B}}=\begin{pmatrix}[1.5]    
\frac{w_1}{v_1}  & \frac{w_2}{v_2}  &  \ldots & \frac{w_n}{v_n} & 0 \\ 
\frac{w_1}{v_1} \, a_1    &  \frac{w_2}{v_2} \, a_2    & \ldots & \frac{w_n}{v_n} \, a_{n} & 0 \\
\vdots &  \vdots &   \ddots & \vdots & \vdots \\
\frac{w_1}{v_1} \, a_1^{\frac{n-k-4}{2}}    & \frac{w_2}{v_2} \, a_2^{\frac{n-k-4}{2}}    & \ldots & \frac{w_n}{v_n} \, a_{n}^{\frac{n-k-4}{2}} & 0 \\
\frac{w_1}{v_1} \, a_1^{\frac{n-k-2}{2}}    & \frac{w_2}{v_2} \, a_2^{\frac{n-k-2}{2}}    & \ldots & \frac{w_n}{v_n} \, a_{n}^{\frac{n-k-2}{2}} & -\eta 
\end{pmatrix}.
\end{align}
Hence, $\mathcal{A}\star \mathcal{B}$ is an $[n+1,n-k]_q$ linear code whose generator matrix is  
\begin{align*}
G_{\mathcal{A}\star \mathcal{B}}=
\begin{pmatrix}[1.5]
\frac{w_1}{v_1}  &  \frac{w_2}{v_2}  &  \ldots & \frac{w_n}{v_n} & 0 \\ 
\frac{w_1}{v_1} \, a_1    & \frac{w_2}{v_2} \, a_2    & \ldots & \frac{w_n}{v_n} \, a_{n} & 0 \\
\vdots &  \vdots &   \ddots & \vdots & \vdots \\
\frac{w_1}{v_1} \, a_1^{n-k-2}    &\frac{w_2}{v_2} \, a_2^{n-k-2}    & \ldots & \frac{w_n}{v_n} \, a_{n}^{n-k-2} & 0 \\
\frac{w_1}{v_1} \, a_1^{n-k-1}    & \frac{w_2}{v_2} \, a_2^{n-k-1}    & \ldots & \frac{w_n}{v_n} \, a_{n}^{n-k-1} & -\eta            
\end{pmatrix}.
\end{align*}
Using (\ref{eq.ETGRS.parity-check matrix}), we get  
$\mathcal{A}\star \mathcal{B}\subseteq \ETGRS_{k}(\SSS,{\bf v},\eta,\infty)^{\perp}$, confirming that the first condition in Definition \ref{def.ECP} is met. By a similar argument as that in the proof of Theorem \ref{th.ECP111}, we verify that the remaining three conditions in Definition \ref{def.ECP} are also satisfied. Thus, 
\[
(\mathcal{A},\mathcal{B})= 
\left(
    \EGRS_{\frac{n-k+2}{2}}(\SSS,{\bf 1},\infty),
    \EGRS_{\frac{n-k}{2}}\left(\SSS,\frac{\www}{\vvv},\infty\right) \diag(\underbrace{1,1,\ldots,1}_n,-\eta)
\right)
\]
is an $\frac{n-k}{2}$-ECP of $\ETGRS_{k}(\SSS,{\bf v},\eta,\infty)$.
\end{IEEEproof}

\begin{remark}
Theorems \ref{th.ECP111} and \ref{th.ECP222} together provide a complete solution to Problem \ref{prob.2} Part 2.
\end{remark}

Having established the $t$-ECPs, we proceed to provide an explicit decoding algorithm for $(+)$-ETGRS codes. Let $(\mathcal{A},\mathcal{B})$ be a $t$-ECP of $\ETGRS_k(\SSS,{\bf v},\eta,\infty)$ and let ${\bf y}={\bf c}+{\bf e}\in \F_q^n$ be a received vector such that ${\bf c}\in \ETGRS_k(\SSS,{\bf v},\eta,\infty)$ and $\wt({\bf e})\leq t$. Based on \cite[Theorem 1]{DK1994}, the first three conditions in Definition \ref{def.ECP} guarantee the existence of a nonzero ${\bf a}\in \mathcal{A}$ such that 
$\langle ({\bf a}\star {\bf b}), {\bf y}\rangle=0$ for any ${\bf b}\in \mathcal{B}$. This allows us to write  
\[
G_{\mathcal{B}}\cdot \diag({\bf y}) \cdot {\bf a}^{\top}={\bf 0}.  
\]
Any solution ${\bf a}\in \mathcal{A}$ of the above system of equations satisfies ${\bf e}\star {\bf a}={\bf 0}$, with $G_{\mathcal{B}}$ being a generator matrix of $\mathcal{B}$ and $\diag({\bf y})$ the $n\times n$ diagonal matrix with the entries of ${\bf y}$ in its main diagonal. Hence, $\supp({\bf e})\subseteq z({\bf a})$, that is, any nonzero solution ${\bf a}$ can locate the possible error positions of ${\bf y}$. 

On the other hand, we know from \cite{P1992} that the fourth condition in Definition \ref{def.ECP} ensures that 
\[
H_{\ETGRS_{k}(\SSS,{\bf v},\eta,\infty)}{\bf x}^{\top} = H_{\ETGRS_{k}(\SSS,{\bf v},\eta,\infty)}{\bf y}^{\top}
=H_{\ETGRS_{k}(\SSS,{\bf v},\eta,\infty)}({\bf c}+{\bf e})^{\top}=H_{\ETGRS_{k}(\SSS,{\bf v},\eta,\infty)}{\bf e}^{\top}
\]
has a unique solution for ${\bf x}=(x_1,x_2,\ldots,x_n)$, 
where $H_{\ETGRS_{k}(\SSS,{\bf v},\eta,\infty)}$ is a parity-check matrix 
of $\ETGRS_{k}(\SSS,{\bf v},\eta,\infty)$ as in (\ref{eq.ETGRS.parity-check matrix}) 
and $x_i = 0$ for any $i \notin z({\bf a})$.

Combining the insights, the received vector ${\bf y}$ is decoded to the codeword 
${\bf c} = {\bf y} - {\bf x}$. Using both Theorems \ref{th.ECP111} and \ref{th.ECP222}, we are ready to give an explicit decoding algorithm for $\ETGRS_k(\SSS, {\bf v}, \infty)$ based on its $t$-ECP.

\begin{theorem}\label{th.decoding ECP}
Algorithm \ref{alg.1} decodes any $(+)$-ETGRS code with parameters $[n+1,k]_q$ in $O(n^3)$ time. 
\end{theorem}
\begin{IEEEproof}
The decoding correctness follows from Theorems \ref{th.ECP111} and \ref{th.ECP222} given the details provided in Algorithm \ref{alg.1}. By \cite[Remark 2.2(1)]{P1996}, the algorithms takes $O((nN)^3)$ time, with $N$ denoting the degree of the field extension of the finite field that contains the $t$-ECP $(\mathcal{A},\mathcal{B})$ over the base field of the linear code of length $n$. 
By Theorems \ref{th.ECP111} and \ref{th.ECP222}, we know that $N=1$, leading to time complexity $O(n^3)$. 
\end{IEEEproof}

The next three remarks compare the time complexity of Algorithm \ref{alg.1} with known ones in the literature. 

\begin{remark}\label{rem.complexity_comparison111}
Beelen \textit{et al.} proposed a decoding method for general TGRS and ETGRS code in \cite{BPN2017} based on the well-known decoding algorithm of GRS and EGRS codes. Sun \textit{et al.} in \cite[Page 1023]{SYJL2025} determined that the time complexity for this decoding method is at least $O(q \, n^2)$. Since the length of an $[n+1,k]_q$ $(+)$-ETGRS code is bounded by the field size $q$, 
Algorithm \ref{alg.1} has the same time complexity as the method in \cite{BPN2017} whenever $O(n)=O(q)$ with better performance when $O(n)=O\left(\sqrt[\ell]{q}\right)$ with $\ell \geq 2$.
\end{remark}

\begin{remark}\label{rem.complexity_comparison222}
Linear codes deployed in practical applications usually have special algebraic properties, such as being Euclidean self-orthogonal. Closely related to our studies here, Zhu and Liao constructed three classes of Euclidean self-orthogonal $(+)$-ETGRS codes with parameters listed in Table \ref{tab.SO_ETGRS}. The lengths of the codes in the first and third classes are at most $O(q)$. The lengths of the codes in the second class is at most $O(\sqrt{q})$. Hence, the respective computational complexities of Beelen’s decoding algorithm for these three classes 
are at least $O(n^3)$, $O(n^4)$, and $O(n^3)$, as shown in the last column of Table \ref{tab.SO_ETGRS}. Algorithm \ref{alg.1} performs better for the second class of codes while being at least on par for the other two classes. In special cases, see the treatment in \cite{DK1994} for examples, the complexity of Algorithm \ref{alg.decoding} can be reduced to $O(n^2)$.

\begin{table*}[ht!]
\centering
\caption{{Three classes of Euclidean self-orthogonal $(+)$-ETGRS codes from \cite{ZL2024}} and their decoding complexities in \cite{BPN2017}.}\label{tab.SO_ETGRS}       
\centering
\setlength{\tabcolsep}{8pt} 
\renewcommand{\arraystretch}{1.2} 
\begin{tabular}{c|l|l|l|c}
\toprule
No. & Parameters  & Conditions & Reference & Decoding Complexity in \cite{BPN2017}\\ 
\midrule
1 & $[2k+1,k]_{2^m}$ & $3\leq k\leq 2^{m-1}-1$  & \cite[Cor.~4.9]{ZL2024} & $O(n^3)$ \\ 
2 & $[2k+1,\ell]_{p^{2m}}$ & $3\leq k\leq \frac{p^m-1}{2}$, $p$ is odd, $3\leq \ell\leq k-1$  & \cite[Cor.~4.10]{ZL2024} & $O(n^4)$ \\
3 & $[p^{m}-p^{m-r}+1,\ell]_{p^{m}}$ & $m=2 \, r$,~$3\leq \ell\leq \frac{p^m-p^{m-r}}{2}-1$, $p$ is odd & \cite[Cor.~4.11]{ZL2024} & $O(n^3)$ \\ 
\bottomrule
\end{tabular}
\end{table*} 
\end{remark}

\begin{remark}\label{rem.complexity_comparison333}
A decoding strategy for general TGRS codes, which is often faster than the one in \cite{BPN2017} was proposed in \cite{BPR2022IT}. 
These two works were done by the same authors. 
They mentioned in \cite[Page 3056]{BPR2022IT} that they were unable to rigorously 
prove that the decoding procedure would succeed for all error vectors up to the maximal decoding radius. Jia \textit{et al.} in~\cite{JYS2025}, Sui and Yue in~\cite{SY2023}, and Sun \textit{et al.} in~\cite{SYJL2025} used the Berlekamp–Massey algorithm for GRS codes and the extended Euclidean algorithm to design efficient decoding algorithms for the duals of $(+)$-TGRS codes. The Euclidean dual of a $(+)$-TGRS code, however, is not necessarily a $(+)$-TGRS code, see, \textit{e.g.}, the discussion in \cite{HYN2021DCC} and \cite{JYS2025} for further details. Whether their decoding algorithms would work directly on $(+)$-ETGRS codes was left untreated.
\end{remark}

\begin{algorithm}[!t]\label{alg.decoding}
\SetKwData{Left}{left}\SetKwData{This}{this}\SetKwData{Up}{up}
\SetKwFunction{Union}{Union}\SetKwFunction{FindCompress}{FindCompress}
\SetKwInOut{Input}{input}\SetKwInOut{Output}{output}
\caption{A Decoding Algorithm for $\ETGRS_k(\SSS,{\bf v},\eta,\infty)$ in terms of $t$-ECPs}\label{alg.1}
\KwIn{A received vector ${\bf y}=(y_1,y_2,\ldots,y_{n+1})\in \F_q^{n+1}$}
\KwOut{A decoded codeword ${\bf c}\in \ETGRS_k(\SSS,{\bf v},\eta,\infty)$ or a failure message}
\Begin{
    $H_{\ETGRS_k(\SSS,{\bf v},\eta,\infty)} \gets$ the parity-check matrix of 
    $\ETGRS_k(\SSS,{\bf v},\eta,\infty)$ as in \eqref{eq.ETGRS.parity-check matrix}\\
    \uIf {$H_{\ETGRS_k(\SSS,{\bf v},\eta,\infty)}\cdot {\bf y}^{\top}={\bf 0}$} 
        {${\bf c} \gets {\bf y}$\\
        \Return ${\bf c}$ is the decoded codeword} 
    \Else {
        $\diag({\bf y}) \gets \diag(y_1,y_2,\ldots,y_{n+1})$\\
        
        \uIf {$n-k$ is odd} 
        {$t \gets \frac{n-k-1}{2}$\\
        $\mathcal{A} \gets \overline{\GRS_{\frac{n-k+1}{2}}(\SSS,{\bf 1})}({\bf 0})$\\
        $\mathcal{B} \gets \EGRS_{\frac{n-k-1}{2}}(\SSS,\frac{\www}{\vvv},\infty)$\\}

        \Else {$t \gets \frac{n-k}{2}$\\
        $\mathcal{A} \gets \EGRS_{\frac{n-k+2}{2}}(\SSS,{\bf 1},\infty)$\\
        $\mathcal{B} \gets \EGRS_{\frac{n-k}{2}}(\SSS,\frac{\www}{\vvv},\infty) \,  \diag(\underbrace{1,1,\ldots,1}_n, -\eta)$\\
        }
        $G_{\mathcal{A}}\gets $ the generator matrix of $\mathcal{A}$ as in \eqref{eq.ECP_GM111} or \eqref{eq.ECP_GM222}\\
        $G_{\mathcal{B}}\gets $ the generator matrix of $\mathcal{B}$ as in \eqref{eq.ECP_GM111} or \eqref{eq.ECP_GM222}\\
        ${\bf u}({\bf y}) \gets \left\{{\bf u}=(u_1,u_2,\ldots,u_{\dim(\mathcal{A})}):~ G_{\mathcal{B}}\cdot \diag({\bf y}) \cdot G_{\mathcal{A}}^{\top}\cdot {\bf u}^{\top}={\bf 0}\right\}$\\

        \uIf {${\bf u}({\bf y})=\{{\bf 0}\}$} 
            {\Return ${\bf y}$ has more than $t$ errors} 
        \Else {
            ${\bf u}'\gets $ any nonzero vector in ${\bf u}({\bf y})$\\
            $Z\gets z({\bf u}' \, G_\mathcal{A})$\\
            ${\bf x}({\bf y})\gets \left\{{\bf x} \, : \,          H_{\ETGRS_k(\SSS,{\bf v},\eta,\infty)} \cdot {\bf x}^{\top} = H_{\ETGRS_k(\SSS,{\bf v},\eta,\infty)}\cdot {\bf y}^{\top} \mbox{ with } x_i=0 \mbox{ for all }i \notin Z\right\}$\\
            \uIf {${\bf x}({\bf y})= \{{\bf x}'\}$ and $\wt({\bf x}')\leq t$} 
                {${\bf c} \gets {\bf y}-{\bf x}'$\\
                \Return ${\bf c}$ is the decoded codeword}
        \Else {
            \Return ${\bf y}$ has more than $t$ errors
            } 
            } 
    }       
}
\end{algorithm}

The next example illustrates the decoding procedure in Algorithm \ref{alg.1}.

\begin{example}\label{exam.decoding}
Let $\vvv={\bf 1}\in (\F_{16}^*)^{11}$ and let $\w$ be a primitive element of $\F_{16}$ such that $\w^4+\w+1=0$. Let
\[ 
\SSS=\{1, \w, \w^{12}, \w^2, \w^{13}, \w^{14}, \w^4, \w^5, \w^6, \w^7, \w^9\}\subseteq \F_{16} \mbox{ and } \eta=\w^6.
\]
Letting ${\bf w}=(\w^3, \w^4, \w^4, \w^6, \w^7, \w^{11}, \w, \w^8, 1, \w^6, \w^7)$, we build the $(+)$-ETGRS code $\ETGRS_5(\SSS,{\bf 1},\w^6,\infty)$ whose generator and parity-check matrices are
\begin{align*}
G_{\ETGRS_5(\SSS,{\bf 1},\w^6,\infty)} &=
\begin{pmatrix}
1 &  1 & 1 & 1 & 1 & 1 & 1 & 1 & 1 &  1 & 1 & 0 \\
1 & \w & \w^{12} & \w^2 & \w^{13} & \w^{14} & \w^4 & \w^5 & \w^6 & \w^{7} & \w^{9} & 0 \\
1 & \w^2 & \w^9 & \w^4 & \w^{11} & \w^{13} & \w^{8} & \w^{10} & \w^{12} & \w^{14} & \w^{3} & 0  \\
1 & \w^3 & \w^6 & \w^6 & \w^{9} & \w^{12} & \w^{12} & 1 & \w^{3} & \w^{6} & \w^{12} & 0 \\
\w^{13} & \w^{13} & \w^2 & \w^{10} & \w^8 & \w^6 & \w^{6} & \w^2 & \w^5 & \w^{4} & 0 & 1
\end{pmatrix} \mbox{ and}\\
H_{\ETGRS_5(\SSS,{\bf 1},\w^6,\infty)} &=
\begin{pmatrix}
\w^3 & \w^4 & \w^4 & \w^6 & \w^7 & \w^{11} & \w & \w^8 &   1 & \w^6 & \w^7 & 0\\
\w^3 & \w^5 &   \w & \w^8 & \w^5 & \w^{10} & \w^5 & \w^{13} & \w^6 & \w^{13} &   \w &   0\\
\w^3 & \w^6 & \w^{13} & \w^{10} & \w^3 & \w^9 & \w^9 & \w^3 & \w^{12} & \w^5 & \w^{10} &   0\\
\w^3 & \w^7 & \w^{10} & \w^{12} &   \w & \w^8 & \w^{13} & \w^8 & \w^3 & \w^{12} & \w^4 &   0\\
\w^3 & \w^8 & \w^7 & \w^{14} & \w^{14} & \w^7 & \w^2 & \w^{13} & \w^9 & \w^4 & \w^{13} &   0\\
\w^3 & \w^9 & \w^4 &   \w & \w^{12} & \w^6 & \w^6 & \w^3 &   1 & \w^{11} & \w^7 & \w^6\\
\w^3 & \w^{10} &   \w & \w^3 & \w^{10} & \w^5 & \w^{10} & \w^8 & \w^6 & \w^3 &  \w & \w^2
\end{pmatrix}.
\end{align*}
Upon receiving
\[
{\bf y}=(\w^{13},\w^{10},\w^2,\w^{10},\w^5,\w^6,\w^7,\w^2,\w^5,\w^4,0,1)\in \F_{16}^{12},
\]
we have
\[
\diag({\bf y})=\diag(\w^{13},\w^{10},\w^2,\w^{10},\w^5,\w^6,\w^7,\w^2,\w^5,\w^4,0,1)
\]
and proceed as follows.
\begin{itemize}
\item Observing that the sum of the elements in $\{\w, \w^{12}, \w^5, \w^7, \w^9\}\subseteq \SSS$ is $-\eta^{-1}=\w^9$, by Lemma \ref{lem.parameters_of_ETGRS} Part 1 we know that $\ETGRS_5(\SSS,{\bf 1},\w^6,\infty)$ is a $[12,5,7]_{16}$ NMDS code. We infer that ${\bf y}$ is not a codeword of $\ETGRS_5(\SSS,{\bf 1},\w^6,\infty)$ since 
\[
H_{\ETGRS_5(\SSS,{\bf 1},\w^6,\infty)}\cdot {\bf y}^{\top}=(\w^{13}, \w, \w^{14}, 1, \w^4, \w^3, \w^6)^{\top}\neq {\bf 0}.
\]

\item Since $2$ divides $6$, Lines 13-15, 17, and 18 of Algorithm \ref{alg.1} yield $t=3$, implying that $(\mathcal{A},\mathcal{B})$ is a $3$-ECP of $\ETGRS_5(\SSS,{\bf 1},\w^6,\infty)$, 
where  
\[
\mathcal{A}=\EGRS_{4}(\SSS,{\bf 1},\infty) \mbox{ and }
\mathcal{B}=\EGRS_{3}(\SSS,\mathbf{w},\infty) \, \diag(\underbrace{1,1,\ldots,1}_n,\w^6),
\]
with 
\begin{align*} 
G_{\mathcal{A}} &=
\begin{pmatrix} 
1 &  1 & 1 & 1 & 1 & 1 & 1 & 1 & 1 &  1 & 1 & 0  \\
1 & \w & \w^{12} & \w^2 & \w^{13} & \w^{14} & \w^4 & \w^5 & \w^6 & \w^{7} & \w^{9} & 0  \\
1 & \w^2 & \w^9 & \w^4 & \w^{11} & \w^{13} & \w^{8} & \w^{10} & \w^{12} & \w^{14} & \w^{3} & 0  \\
1 & \w^3 & \w^6 & \w^6 & \w^{9} & \w^{12} & \w^{12} & 1 & \w^{3} & \w^{6} & \w^{12} & 0 
\end{pmatrix}
\mbox{ and}\\
G_{\mathcal{B}} &=
\begin{pmatrix} 
\w^3 & \w^4 & \w^4 & \w^6 & \w^7 & \w^{11} & \w & \w^8 &   1 & \w^6 & \w^7 & 0\\
\w^3 & \w^5 &   \w & \w^8 & \w^5 & \w^{10} & \w^5 & \w^{13} & \w^6 & \w^{13} &   \w &   0\\
\w^3 & \w^6 & \w^{13} & \w^{10} & \w^3 & \w^9 & \w^9 & \w^3 & \w^{12} & \w^5 & \w^{10} &   \w^6
\end{pmatrix}.
\end{align*}
\item Line 19 of Algorithm \ref{alg.1} helps us solve the systems of linear equations 
\[
G_{\mathcal{B}}\cdot \diag({\bf y}) \cdot G_{\mathcal{A}}^{\top}\cdot {\bf u}^{\top}={\bf 0},
\]
yielding ${\bf u}({\bf y})=\{a(1,\w^4,\w^3,\w^{12}) \, : \, a \in \F_{16}\}$. Taking ${\bf u}'=(1,\w^4,\w^3,\w^{12})$, we get  
\[
{\bf u}' \, G_{\mathcal{A}}=(\w^8,0,\w^2,\w^{11},0,1,0,\w^{14},\w^5,\w^4,\w^9,\w^{12}) \mbox{ and } Z = z \left({\bf u}' \,  G_{\mathcal{A}}\right) = \{2,5,7\}.
\]
\item Based on Line 25 of Algorithm \ref{alg.1}, we let 
${\bf x}=(0,x_2,0,0,x_5,0,x_7,0,0,0,0,0)$ and solve the systems of linear equations  
\[
H_{\ETGRS_5(\SSS,{\bf 1},\w^6,\infty)} \, {\bf x}^{\top}=H_{\ETGRS_5(\SSS,{\bf 1},\w^6,\infty)} \, {\bf y}^{\top}
\]
to get the unique solution ${\bf x}'=\{0,\w^9,0,0,\w^4,0,\w^{10},0,0,0,0,0\}$ whose $\wt({\bf x}')=3$. 
       
\item Finally, using Lines 26 and 27 of Algorithm \ref{alg.1}, the received vector ${\bf y}$ is decoded to the codeword 
\[
{\bf c}={\bf y}-{\bf x}'=(\w^{13},\w^{13},\w^2,\w^{10},\w^8,\w^6,\w^6,\w^2,\w^5,\w^4,0,1).
\]
The decoding process is successful since $H_{\ETGRS_5(\SSS,{\bf 1},\w^6,\infty)}\cdot {\bf c}^{\top}={\bf 0}$.
\end{itemize}
\end{example}

\subsection{Decoding non-GRS MDS Han-Zhang codes by their deep holes}

Bringing our focus back to Problem \ref{prob.3}, we recall useful results on the covering radius and deep holes of linear codes. Further details can be found in \cite{FXZ2025,WDC2023,SHOS2022,SHO2023,SLHO2025,C2025}, and the references therein.

\begin{definition}
For an $[n,k]_q$ code $\C$ and any vector $\mathbf{u}\in \F_q^n$, the {\em error distance} of $\uuu$ to $\C$ is 
\[
d(\mathbf{u},\C)=\min\{d(\mathbf{u},\mathbf{c}):~\mathbf{c}\in \C\}.
\]
We call $\rho(\C)=\max\{d(\mathbf{u}, \C):~\mathbf{u}\in \F_q^n\}$ the {\em covering radius} of $\C$. The {\em deep holes} of $\C$ are the vectors in $\F_q^n$ whose error distances to $\C$ achieve $\rho(\C)$.        
\end{definition}

Wu, Ding, and Chen have established an elegant connection between the MDS property of the second extended code of a given MDS code and the covering radius and deep holes of its Euclidean dual in \cite{WDC2023}. We rephrase it in the next lemma. 

\begin{lemma}{\rm (\!\! \cite[Theorem 6]{WDC2023})}\label{lem.deep holes}
If $\C$ is an $[n,k,n-k+1]_q$ MDS code, then, for any $\uuu\in \F_q^n$, its second extended code $\overline{\C}(\uuu)$ is an MDS code if and only if $\rho(\C^{\perp})=k$ and $\uuu$ is a deep hole of $\C^{\perp}$. 
\end{lemma}
To obtain the covering radius and deep holes of non-GRS MDS Han-Zhang codes using Lemma \ref{lem.deep holes}, 
we start by giving another expression of $(+)$-ETGRS codes in terms of the second extended codes of $(+)$-TGRS codes using the following lemma.

\begin{lemma}{\rm(\!\!\cite[Lemma 2.9]{LSZ2024})}\label{lem.wi}
\label{lem.PRA}
Let $\{a_1,a_2,\ldots,a_n\}\subseteq \F_q$ with $n\geq 3$ be given. If $w_i$ is as in \eqref{eq:wi}, then  
\begin{equation}\label{eq.ui value}
\sum_{i=1}^{n} a_i^{\ell} \, w_i=
\begin{cases}
0, & \mbox{if } 0\leq \ell\leq n-2, \\
1, & \mbox{if } \ell=n-1, \\
\sum_{i=1}^n a_i, & \mbox{if } \ell=n.
\end{cases}
\end{equation}
\end{lemma}

\begin{theorem}\label{th.ETGRS is second kind of extended code}
Let $t\in\{n-k-1,n-k\}$ and let
\begin{equation}\label{eq:s}
s=\begin{cases}
\eta^{-1}, & \mbox{if } t=n-k-1, \\
\left(1+ \eta \, \sum_{i=1}^{n}a_i\right)^{-1}, &  \mbox{if } t=n-k \mbox{ and } 1+\eta \, \sum_{i=1}^{n}a_i\neq 0.
\end{cases}
\end{equation}
If $\uuu=(u_1,u_2,\ldots,u_{n})\in \F_q^{n}$ is such that 
\begin{equation}\label{eq:u}
u_i=\frac{s \, a_i^t}{v_i \, \prod_{1\leq j\neq i\leq n}(a_i-a_j)} \mbox{ for } 1\leq i\leq n,
\end{equation}
then 
\[
\ETGRS_k(\SSS,\mathbf{v},\eta,\infty)=\overline{\TGRS_k(\SSS,\mathbf{v},\eta)}(\uuu).
\]
\end{theorem}
\begin{IEEEproof}
By \eqref{eq.first kind of extended codes} and Definition \ref{def.extended code 2}, it suffices to prove that 
\[
G_{\TGRS_k(\SSS,\mathbf{v},\eta)} \, \uuu^{\top}= (\underbrace{0,0,\ldots,0}_{k-1},1)^{\top}.
\]
For any $1\leq \ell\leq k-1$, since $2\leq k\leq n -1$ and $0\leq n-k-1\leq t\leq n-k$, we have 
\[
n\geq 3 \mbox{ and } 0\leq\ell-1 \leq t+\ell-1\leq n-k+\ell-1\leq n-2.
\]
By Lemma \ref{lem.wi} and the definition of $w_i$ in \eqref{eq:wi}, we obtain
\begin{equation}\label{eq:aspect1}
\sum_{i=1}^{n} v_i \, a_i^{\ell-1} \, u_i = s \sum_{i=1}^{n}a_i^{t+\ell-1} \, w_i=0
\end{equation}
and observe that
\begin{multline}\label{eq:aspect2}
\left(v_1 \, \left(a_1^{k-1} + \eta \, a_1^k \right), v_2 \, \left(a_2^{k-1}+\eta \, a_2^k \right), \ldots, v_n \, \left(a_n^{k-1}+\eta \, a_n^k \right)\right) \, \uuu^{\top} 
= \sum_{i=1}^{n} v_i \, a_i^{k-1} \, u_i+ \sum_{i=1}^{n} \eta \, v_i \, a_i^k \, u_i \\
= s \, \sum_{i=1}^{n}a_i^{t+k-1} \, w_i+ s \, \eta \, \sum_{i=1}^{n} a_i^{t+k} \, w_i 
= \begin{cases} 
0 +s \, \eta, & \mbox{if } t=n-k-1, \\
s+ s \, \eta \, \sum_{i=1}^{n} a_i, & \mbox{if } t=n-k,
\end{cases} \\
= 1 \mbox{ if } t=n-k-1, \mbox{ or } t=n-k \mbox{ and } 
1+\eta \, \sum_{i=1}^{n}a_i\neq 0. 
\end{multline}
Combining \eqref{eq:aspect1} and \eqref{eq:aspect2} completes the proof. 
\end{IEEEproof}
 
Based on Lemma \ref{lem.deep holes} and Theorem \ref{th.ETGRS is second kind of extended code}, in certain cases we have useful information about the covering radius and deep holes of $\TGRS_k(\SSS,\mathbf{v},\eta)^{\perp}$ to successfully perform maximum-likelihood decoding.

\begin{theorem}\label{th.covering radius and deep holes}
If $-\eta^{-1} \notin \left\{ \sum_{a_i\in I}a_i  \, : \, \mbox{ for all } I\subseteq \SSS \mbox{ such that } |I|=k \right\}$, 
then 
\[
\rho\left(\TGRS_k(\SSS,\mathbf{v},\eta)^{\perp}\right)=k.
\]
Let $a\in \F_q^*$ and let $\ccc^{\perp}\in \TGRS_k(\SSS,\mathbf{v},\eta)^{\perp}$. If $\uuu$ is the vector defined in \eqref{eq:u} based on the $s$ in \eqref{eq:s}, then any vector of the form $a \, \uuu+\ccc^{\perp}$ is a deep hole of $\TGRS_k(\SSS,\mathbf{v},\eta)^{\perp}$.
\end{theorem}
\begin{IEEEproof}
Let us recall that $\SSS = \{a_1,a_2,\ldots,a_n\}$. By Lemma \ref{lem.parameters_of_ETGRS} Part 1, under the given conditions, $\TGRS_k(\SSS,\mathbf{v},\eta)$ is an $[n,k,n-k+1]_q$ MDS code and $\ETGRS_k(\SSS,\mathbf{v},\eta,\infty)$ is an $[n+1,k,n-k+2]_q$ MDS code. Combining Theorem~\ref{th.ETGRS is second kind of extended code} and Lemma~\ref{lem.deep holes}, we confirm that $\rho(\TGRS_k(\SSS,\mathbf{v},\eta)^{\perp})=k$ and 
$\uuu$ is a deep hole of $\TGRS_k(\SSS,\mathbf{v},\eta)^{\perp}$. Since $a\in \F_q^*$ and $\ccc^{\perp}\in \TGRS_k(\SSS,\mathbf{v},\eta)^{\perp}$, we can verify that 
\[
d\left(a\uuu+\ccc^{\perp},\TGRS_k(\SSS,\mathbf{v},\eta)^{\perp}\right)=d\left(\uuu,\TGRS_k(\SSS,\mathbf{v},\eta)^{\perp}\right).
\]
Thus, all vectors of the form $a \, \uuu+\ccc^{\perp}$ are deep holes of $\TGRS_k(\SSS,\mathbf{v},\eta)^{\perp}$.
\end{IEEEproof}

We can separate the deep holes $a \, \uuu+\ccc^{\perp}$ shown in Theorem \ref{th.covering radius and deep holes} into two classes.
\begin{itemize}
\item {\bf Class 1:} The deep hole $a \, \uuu+\ccc^{\perp}$ is such that $a\in \F_q^*$, $\ccc^{\perp}\in \TGRS_k(\SSS,\mathbf{v},\eta)^{\perp}$, and $\uuu:=(u_1,u_2,\ldots,u_n)$ satisfies 
\[
u_i=\frac{\eta^{-1} \, a_i^{n-k-1}}{v_i \, \prod_{1\leq j \neq i \leq n} (a_i-a_j)} \mbox{ for } 1\leq i\leq n.
\]
\item {\bf Class 2:} The deep hole $a \, \uuu+\ccc^{\perp}$ is such that $a\in \F_q^*$, $\ccc^{\perp}\in \TGRS_k(\SSS,\mathbf{v},\eta)^{\perp}$, and $\uuu:=(u_1,u_2,\ldots,u_n)$ satisfies
\[
1+ \eta \, \sum_{i=1}^{n}a_i\neq 0 \mbox{ and } 
u_i=\frac{(1+\eta \, \sum_{i=1}^{n}a_i)^{-1} \, a_i^{n-k}} 
{v_i \, \prod_{1\leq j\neq i\leq n}(a_i-a_j)} \mbox{ for } 1\leq i\leq n.
\]
\end{itemize}

In \cite[Theorem 2.4]{HYN2021DCC}, Huang \textit{et al.} determined the parity-check matrix of $\TGRS_k(\SSS,\mathbf{v},\eta)$ by considering three distinct cases. Table~\ref{tab:deep_hole} summarizes the result of Theorem~\ref{th.covering radius and deep holes} by viewing the parity-check matrix of a linear code as a generator matrix of its Euclidean dual. 

\begin{table*}[ht!]
\centering
\caption{Deep holes of $\TGRS_k(\SSS,\mathbf{v},\eta)^{\perp}$ for $-\eta^{-1} \notin \left\{ \sum_{a_i\in I} a_i  \, : \, \mbox{ for all } I \subseteq \SSS \mbox{ such that } |I|=k \right\}$. The class of linear codes $\C(\SSS,n-k,\frac{\www}{\vvv})$ from \cite{HZ2024} is denoted by ${\rm HZ}_{n-k}\left(\SSS,\frac{\www}{\vvv}\right)$. We use “\ding{51}” to indicate that the code has deep holes in the class and “\ding{55}” to indicate that it does not.}\label{tab:deep_hole} 
\setlength{\tabcolsep}{8pt} 
\renewcommand{\arraystretch}{1.3} 
\begin{tabular}{l|l|c|c}
\toprule
Conditions  & $\TGRS_k(\SSS,\mathbf{v},\eta)^{\perp}$ & Class 1 & Class 2\\ 
\midrule
$\sum_{i=1}^{n}a_i\neq 0$ and $1+\eta \sum_{i=1}^{n}a_i= 0$  & ${\rm HZ}_{n-k}\left(\SSS,\frac{\www}{\vvv}\right)$ & \ding{51} & \ding{55}\\
$\sum_{i=1}^{n}a_i\neq 0$ and $1+\eta \sum_{i=1}^{n}a_i\neq 0$  & 
$\TGRS_{n-k}\left(\SSS,\frac{\www}{\vvv},\frac{\eta}{1+\eta \sum_{i=1}^{n}a_i}\right)$  & \ding{51} & \ding{51} \\ 
$\sum_{i=1}^{n}a_i= 0$ & $\TGRS_{n-k}\left(\SSS,\frac{\www}{\vvv},-\eta\right)$  & \ding{51} & \ding{51} \\ 
\bottomrule
\end{tabular}
\end{table*} 

When $\TGRS_k(\SSS,\mathbf{v},\eta)^{\perp}$ is a Han-Zhang code ${\rm HZ}_{n-k}\left(\SSS,\frac{\www}{\vvv}\right)$, Table \ref{tab:deep_hole} provides its covering radius along with a class of deep holes, \emph{which have not been previously identified}. Since the condition $1+\eta \sum_{i=1}^{n}a_i= 0$ implies that $2+\eta \sum_{i=1}^{n}a_i\neq 0$, we can conclude by Theorem \ref{th.non-GRS TGRS for k=n/2111} that ${\rm HZ}_{n-k}\left(\SSS,\frac{\www}{\vvv}\right)$ is non-GRS MDS for any $3\leq k\leq n-3$. Thus, Problem \ref{prob.3} is completely solved.

On the occasion of $\TGRS_k(\SSS,\mathbf{v},\eta)^{\perp}$ being in fact a $(+)$-TGRS code, one can further verify from the structure of $\ccc^{\perp}$ that the two classes of deep holes coincide. Moreover, when $\vvv$ is the all-one vector ${\bf 1}$, Fang \textit{et al.} determined the covering radius $\rho(\TGRS_k(\SSS,{\bf 1}))$ and found a standard class of deep holes of $\TGRS_k(\SSS,{\bf 1})$ in \cite[Theorem 5]{FXZ2025} by calculating the determinants of certain matrices. The deep holes have the same form of ours. Table \ref{tab:deep_hole} shows that we have a different method to obtain the covering radii and deep holes for more general $(+)$-TGRS codes. 

\begin{remark}\label{rem.deep_hole} 
Jin \textit{et al.} have shown in \cite[Proposition \Rmnum{5}.1]{JMXZ2024} that $[n,k,n-k+1]_q$ Han-Zhang codes for $3\leq k\leq \frac{n}{2}$ are non-GRS. Our work here extends the treatment since the dual of a Han-Zhang code is not necessarily a Han–Zhang code.
\end{remark}

\begin{example}\label{exam.deep_holes}
Let $\SSS=\{2, 4, 5, 8, 9, 10\}\subseteq \F_{13}$. Let $\vvv=\www$, with $w_i$ as in \eqref{eq:wi}, and let $\eta=1$. We verify that
\[
12=-\eta^{-1}\notin \left\{\sum_{a_i\in I}a_i  \, : \, \mbox{ for all } I\subseteq \SSS \mbox{ such that } |I|=3\right\}, ~  \sum_{i=1}^{6}a_i=12\neq 0 \mbox{, and } 1+\eta\sum_{i=1}^{6}a_i=0.
\]
By Theorem \ref{th.covering radius and deep holes} and Remark \ref{rem.deep_hole}.1) the code  ${\rm HZ}_3(\SSS,{\bf 1})$ is a $[6,3,4]_{13}$ non-GRS MDS code with covering radius $3$ and 
a family of deep holes 
\[
\left\{ a \, (4,3,12,12,3,9)+\ccc \, : \,  
a \in \F_{13}^* \mbox{ and } \ccc\in {\rm HZ}_3(\SSS,{\bf 1}) \right\}.
\]
One can check the correctness of the claims by \texttt{MAGMA} \cite{magma}.  
\end{example}

\section{Conclusions}\label{sec.conclusion}

Each code from the classes of $(+)$-TGRS and $(+)$-ETGRS codes is either a non-GRS MDS code or an NMDS code. Their suitability for applications in communication, cryptography, and storage systems have recently been more actively explored. Many of their properties have been characterized in the literature. There were gaps regarding their non-GRS MDS properties and open problems on their decoding properties related to ECPs and deep holes. Problems \ref{prob.1}, \ref{prob.2}, and \ref{prob.3} summarize particular gaps that we have investigated in this work. 

Theorems \ref{th.non-GRS TGRS for k=n/2111} and \ref{th.non-GRS TGRS for k=n/2222} identify three sets of sufficient conditions under which a given $(+)$-TGRS code with parameters $\left[n,\frac{n}{2}\right]_q$ is non-GRS. We prove that any $(+)$-ETGRS code with parameters $\left[n+1,\frac{n+1}{2}\right]_q$ is non-GRS in Theorem \ref{th.non-GRS ETGRS} via the Schur square of its dual as characterized in Theorem \ref{th.Schur square of ETGRS}. Applying these main results, we derive criteria for the existence of Hermitian self-dual $(+)$-TGRS and $(+)$-ETGRS codes in Corollaries \ref{coro.HSD_TGRS111}, \ref{coro.HSD_TGRS222}, and \ref{cor.no_galois_sd_codes}. 

Switching our focus to decoding in terms of their ECPs, we provide the nonexistence result on the $\ell$-ECPs of $(+)$-TGRS codes with parameters $\left[n,\frac{n}{2},\frac{n}{2}+1\right]_q$ whenever $n = 4 \, \ell$ in Theorem \ref{th.no_ECP_TGRS}. We characterize the explicit forms of $(+)$-ETGRS codes with parameters $[n+1,k]_q$ and use them to confirm the existence of $\frac{n-k-1}{2}$-ECPs and $\frac{n-k}{2}$-ECPs based on the parity of $n-k$ in Theorems \ref{th.ECP111} and \ref{th.ECP222}. We propose a decoding algorithm for $(+)$-ETGRS codes in Theorem \ref{th.decoding ECP} and Algorithm \ref{alg.decoding} and identify cases for which the algorithm performs better than prior methods in Remarks \ref{rem.complexity_comparison111} and \ref{rem.complexity_comparison222}. 

In exploring decoding properties related to deep holes, we establish a new connection between $(+)$-TGRS and $(+)$-ETGRS codes in Theorem \ref{th.ETGRS is second kind of extended code}, and utilize this connection to determine the covering radius and deep holes of non-GRS MDS Han-Zhang codes in Theorem \ref{th.covering radius and deep holes}. Taken together, our results provide partial or complete solutions to Problems \ref{prob.1}, \ref{prob.2}, and \ref{prob.3}. Numerous examples throughout the paper illustrate our findings.

We end by mentioning two directions worth exploring further. The first one is to investigate the non-GRS properties of $(+)$-TGRS codes with parameters $\left[n, \frac{n}{2}, \frac{n}{2} + 1\right]_q$ when $2 + \eta \sum_{i=1}^{n} a_i = 0$. The second one is to explore new approaches for determining the covering radius and deep holes of Han-Zhang codes in more general settings.


\end{document}